\documentclass[preprint,showpacs,amsmath,amssymb,floatfix]{revtex4}
\usepackage{amsfonts}
\usepackage{amssymb}
\usepackage{amsmath}
\usepackage{epsfig}
\usepackage{hyperref}
\usepackage{bm}
\begin{document}
\title{Lift and drag in intruders moving through hydrostatic granular media at high speeds}
\author{Fabricio Q. Potiguar}
\email{fqpotiguar@ufpa.br}
\affiliation{Departamento de F\'\i sica, ICEN, Av. Augusto Correa, 1, Guam\'a, 66075-110, Bel\'em, Par\'a, Brazil}
\author{Yang Ding}
\affiliation{Aerospace and Mechanical Engineering, University of Southern California, Los Angeles, \\California 90089, USA}
\begin{abstract}
Recently, experiments showed that forces on intruders dragged horizontally through dense, hydrostatic granular packings mainly depend on the local surface orientation and can be seen as the sum of the forces exerted on small surface elements.
In order to understand such forces more deeply, we perform 2D soft-sphere 
molecular dynamics simulation, on similar set up, of an intruder dragged 
through a $50-50$ 
bi-disperse granular packing, with diameters $0.30$ and $0.34$ cm. We measure, 
for both circular and half-circle shapes, the forces parallel (drag) and 
perpendicular 
(lift) to the drag direction as functions of the drag speed, with $V=10.3-309$ 
cm/s, and intruder depths, with $D=3.75-37.5$ cm. The 
drag forces on an intruder monotonically increase with $V$ and $D$, 
and are larger for the circle. However, the lift force does not depend 
monotonically on $V$ and $D$, and this 
relationship is affected by the shape of the intruder. The vertical 
force was negative for the half-circle, but for a small range of $V$ and $D$, 
we measure positive lift. We find no sign change for the lift on the circle, 
which is always positive. The explanation 
for the nonmonotonic dependence is related to the decrease in contacts on the 
intruder as $V$ increases. This is qualitatively 
similar to supersonic flow detachment from an obstacle. The detachment 
picture is supported by simulation measurements of the velocity field around 
the intruder and force profiles measured on its surface.
\end{abstract}
\maketitle

\section{Introduction}
Granular matter is a generic name given to a system composed of macroscopic, 
athermal particles that have mutual repulsive, dissipative interactions 
\cite{Jaeger96}. It is an intensely studied field in the Physics community 
given the several distinct behaviors shown by such systems as a consequence of 
different external conditions imposed on them.

One of such conditions is that which imposes a flow of particles, named 
granular flow \cite{Wieghardt75,Campbell90,Gold03,Midi04}. Within the several 
granular flow 
examples, the flow around immersed obstacles has received some attention 
lately \cite{Amar01,Rericha02,Boudet08}. One of the objectives of such 
investigations is to measure the force in the obstacle due to interactions 
with the flowing grains, the so called granular drag \cite{Buchholtz98,Albert99,Chehata03,Wass03,Ciamarra04,Soller06}, and lift \cite{Wieg74,Soller06,Ding10,Potiguar11}.

The drag was studied for several distinct situations: for bodies immersed in 
slow, dense flows \cite{Albert99,Chehata03,Soller06}, it was observed that the 
drag is proportional to the intruder size (cylinder diameter or vane width), 
to the squared insertion depth and independent on the mean 
flow speed; for obstacle within fast, dilute flows \cite{Buchholtz98,Wass03}, 
this force was seen to obey the familiar drag law for flows around spheres in 
fast viscous flows: proportional to obstacle diameter and to the square of flow 
speed; finally, studies on impacting bodies \cite{Tsimring05,Katsuragi07,Goldman08} show that the drag is a sum of gravity, Coulomb friction (depth dependent) and inertial drag (proportional to the square of the penetration speed). This is the granular analog of Poncelet law used in ballistics.

The lift was explored for partially submerged vanes \cite{Wieg74,Soller06}, 
where it was seen to scale with several geometrical parameters of the vane. 
For horizontally translating objects \cite{Ding10}, the lift was seen to depend 
mainly on the local surface geometry. The lift on a cylinder and a vertical 
plate was positive, while that on a half-cylinder (with the flat surface facing 
down) was negative. A calculation based on the forces exerted on small 
surface elements was able to nicely predict the value observed in the 
experiments and simulation. Finally, the lift was also investigated in an 
ellipse immersed in a fast, dilute flow \cite{Potiguar11} and was seen to 
vanish for any symmetrical 
orientation of the obstacle related to the flow direction. In addition, the 
force calculated for an ideal gas flow of inelastic particles qualitatively 
reproduced the numerical results, but predicted a smaller force than the one 
measured in the simulations, since the obstacle is shielded from the incoming 
flow by the formation of a shock wave.

This work has the main objective to investigate the behavior of the drag and 
lift forces in a body horizontally translated through a dense granular at 
constant, high speed. In \cite{Ding10}, this force was investigated for a few 
parameters. Here, we present 
simulations that extend the parameter ranges used earlier to include values 
that are not currently accessible to experiments. More specifically, 
we studied the drag and lift forces on an intruder as a function of the drag 
speed, the intruder depth and shape. We note that ours is one of the few (if 
not the only one) studies of forces in intruders in fast, dense conditions.

Our results point to very interesting behaviors of the lift force, which are 
nonmonotonic 
dependence on the drag speed and depth. Essentially, we saw that as the 
drag speed increases, a supersonic granular flow \cite{Buchholtz98,Amar01,Rericha02,Chehata03,Wass03,Amar06,Boudet08,Boudet09,Boudet10} sets in and, as 
will be detailed ahead, deeply affects the lift force. Surprisingly, the drag 
force in this regime (even during the slow transition), increases quadratically 
with the speed and linearly with depth. This is in contrast to a previous 
supersonic drag measurements, in which this force was seen to be 
independent on the flow speed \cite{Chehata03}. On the other hand, it agrees 
with the results of supersonic dilute flows of \cite{Buchholtz98,Wass03}. 
Aside from the total force 
measurements, we also measured flow fields (velocity and number density) as 
well as force profiles along the intruder surface in order to support our 
understanding of the results.

Section \ref{sim} presents the numerical model and parameters used. 
Section \ref{results} has all our drag and lift results. These are followed by 
the fields and profiles data in section \ref{fields}. Finally, we present 
our conclusions in \ref{conclusions}.

\section{Simulation design\label{sim}}
The system has $N=31276$ soft disks (for speed studies) and $N=46900$ (for 
depth studies) in a $50-50$ bi-disperse mixture with diameters and masses 
$d_1=0.30$ cm, $m_1=34.9$ mg and $d_2=0.34$ cm, $m_2=50.8$ mg. The system is 
located in a container of length $L_X=100$ cm. 

The force between two contacting disks is 
given by the model used in \cite{Ding10}. The normal component is:
\begin{equation}
\label{norm_force}
{\bf F}_{ij}^N={\bf f}_{ij}+{\bf f}_{ij}^d.
\end{equation}
The first term is a conservative part, given by the Hertz law:
\[
{\bf f}_{ij}=\kappa\delta^{3/2}{\bf{\hat r}}_{ij},
\]
where $\kappa=1.04\times10^5$ N/m$^{3/2}$ is the hardness, $d_i$ is the $i$-th 
disk diameter, $r_{ij}$ is the distance between the two disks, 
$\delta=(d_i+d_j)/2-r_{ij}$ is the overlap distance and ${\bf{\hat r}}_{ij}$ is 
an unit vector along the normal between the disks' centers. The second term is 
a dissipative, velocity dependent force, given by
\[
{\bf f}_{ij}^d=-\sigma({\bf {\hat r}}_{ij}\cdot {\bf v}_{ij})\delta^{1/2}{\bf{\hat r}}_{ij},
\]
where $\sigma=7.28\times10^{-2}$ N$\cdot$s/m$^{3/2}$ is the normal damping 
coefficient and ${\bf v}_{ij}={\bf v}_i-{\bf v}_j$ is the relative velocity 
between the disks. The tangential force at the contact point is given by the 
following expression: 
\begin{equation}
\label{tang_force}
{\bf F}_{ij}^S=-\left(\mu_{PP}F_N\right){\bf{\hat v}}_{ij}^S,
\end{equation}
where $\mu_{PP}=0.10$, the static friction coefficient between particles (for 
particle-intruder contacts, $\mu_{PI}=0.27$), 
$F_N=\left|{\bf F}_{ij}^N\right|$, and ${\bf{\hat v}}_{ij}^S$ is the unit 
vector along the direction of the relative velocity at the contact point. This 
vector is calculated as: 
\[
{\bf v}_{ij}^S={\bf v}_{ij}-({\bf {\hat r}}_{ij}\cdot {\bf v}_{ij}){\bf{\hat r}}_{ij}-\left(\frac{d_i{\bm{\omega}}_i+d_j{\bm\omega}_j}{d_i+d_j}\right)\times{\bf r}_{ij},
\]
where ${\bm\omega}_i$ is the $i$-th disk angular velocity and 
$v_{ij}^S=\left|{\bf v}_{ij}^S\right|$. The total contact force 
${\bf F}_{ij}={\bf F}_{ij}^N+{\bf F}_{ij}^S$ vanishes if the disks are not in 
contact, i.e., if $\delta<0$. Finally, each grain suffers the 
effect of gravity, given by a vertical force 
\begin{equation}
\label{gravity}
{\bf g}_i=-m_ig{\bf j},
\end{equation}
where $g=981$ cm/s$^2$ is the acceleration of gravity, and ${\bf j}$ is a 
vertical unit vector.

\begin{figure}[h]
\rotatebox{0}{\epsfig{file=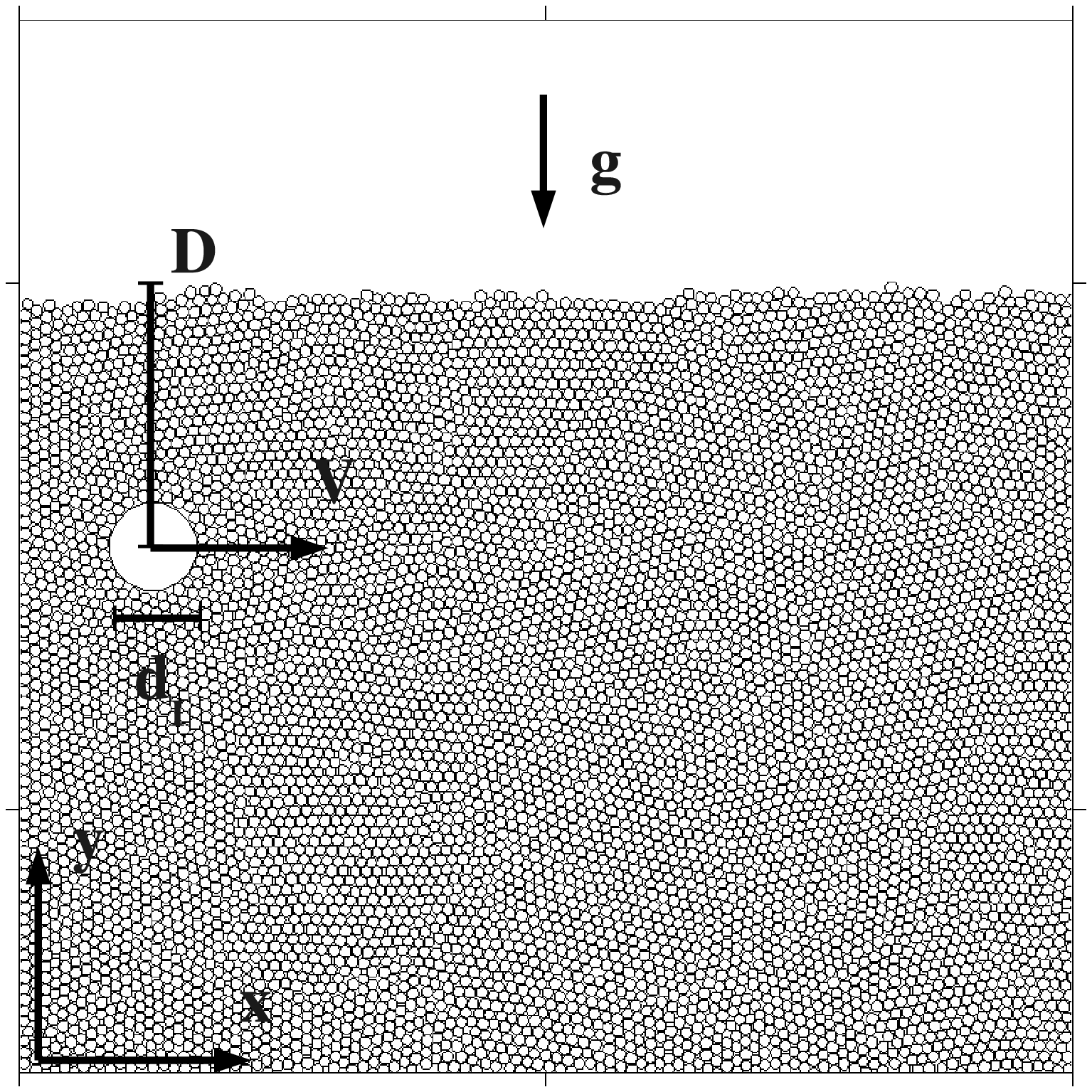,width=13.0cm,height=13.0cm}}
\caption{Pictorial representation of simulation parameters: intruder, of 
diameter $d_I$, is dragged at contant depth, $D$, and speed, $V$, under the 
influence of gravity. The average drag, $\left<R\right>$, is in the $-{\bf V}$ 
direction, while the average lift, $\left<L\right>$, is in either $\pm {\bf g}$ 
direction.
\label{sys_draw}}
\end{figure}

The intruder is either a circle or a half-circle (both with the curved section facing upwards or downwards, which we call inverted half-circle) with a 
diameter $d_I=2.54$ cm. 
It is located, initially, at the position $(x_I=1.5d_I,y_I=H)$, $H$ is the 
vertical position of the intruder as measured from the bottom of the 
container. This parameter is 
related to the depth $D$, vertical position measured with respect to the 
packing free surface, by $D=30.0-H$, since for the number of disks and their 
sizes used here the free surface height, after settling, is $\approx30.0$ cm 
(for depth studies, the free surface is $\approx45.0$ cm tall), see fig. 
\ref{sys_draw}. 
The interactions between the disks and the intruders are the same as those 
given above for two disks. 

At the beginning, all disks are randomly generated without overlap among them 
and with the intruder. At this stage, all intruders are modeled as circles with 
diameter $d_I$ since this facilitates the overlap check. Then, all disks are 
allowed to settle under gravity long enough to bring the packing to rest, 
i.e., when the total kinetic energy per disk is $\approx10^{-4}$ J. 

\begin{figure}[h]
\rotatebox{0}{\epsfig{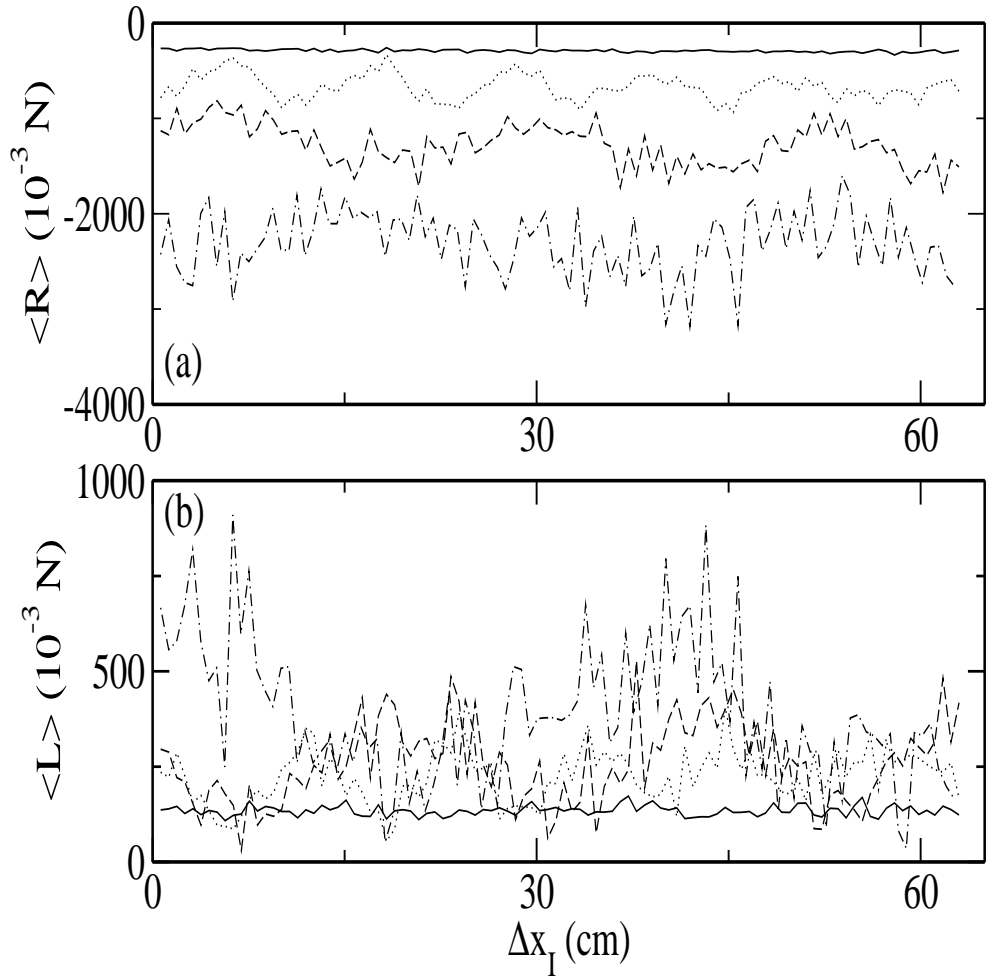}}
\caption{Drag, (a), and lift, (b), vs intruder displacement for depth $3.75$ cm 
and speeds (in cm/s): $10.3$ (solid), $103$ (dotted), $206$ (dashed) and $309$ 
(dot-dashed). 
\label{force_disp}}
\end{figure}

After the settling phase is finished, the intruder is dragged horizontally, 
with a constant speed of $V=10.3-309$ cm/s, with a total of $16$ speeds. We 
also performed measurements at distinct depths, $D=3.75-37.5$ cm, probing 
$10$ distinct depths. Our interest 
is in the drag, $R$, and lift, $L$, forces, respectively, on the intruders as 
well as flow fields and profiles. The first force is in the negative horizontal 
direction (opposite to the intruder displacement) while the second is in the 
vertical direction (perpendicular to the intruder displacement) and can be 
either positive or negative. 
All quantities are presented as averages over time and $5$ independent runs. 
See fig. \ref{force_disp} for time series examples of such measurements.

Our results can be cast as a function of the Froude number:
\[
\mathrm{Fr}=\frac{V^2}{gd_I}.
\]
Given our parameters, this number starts at $4.26\times10^{-2}$ and ends at 
$38.3$, spanning four decades in magnitude. As a comparison, the impact 
experiments of \cite{Walsh03} were performed in the range 
$3.13<\mathrm{Fr}<143$. Notice, then, that our speed range, although 
large, does not include very high Froude numbers. Nevertheless, we will refer 
later in the text to results at low and high speeds. These adjectives mean 
that we are considering, respectively, speed values that are in the lower and 
upper end of our range.

Equations of motion are integrated with a leapfrog scheme 
\cite{rapaport}, with a time step of $0.001$. The settling 
phase is, typically, $10^5$ cycles long, while the drag phase is long 
enough to displace the intruder by $\approx270d_1$.




\section{Numerical Results\label{results}}
The next two subsections have the numerical results for the drag and lift 
forces on the intruders as functions of the drag speed and depth. Also, we 
give a few remarks regarding their interpretation 
that will be supported with the data for the fields and profiles.

\subsection{Drag force\label{drag}}

\begin{figure}[h]
\rotatebox{0}{\epsfig{file=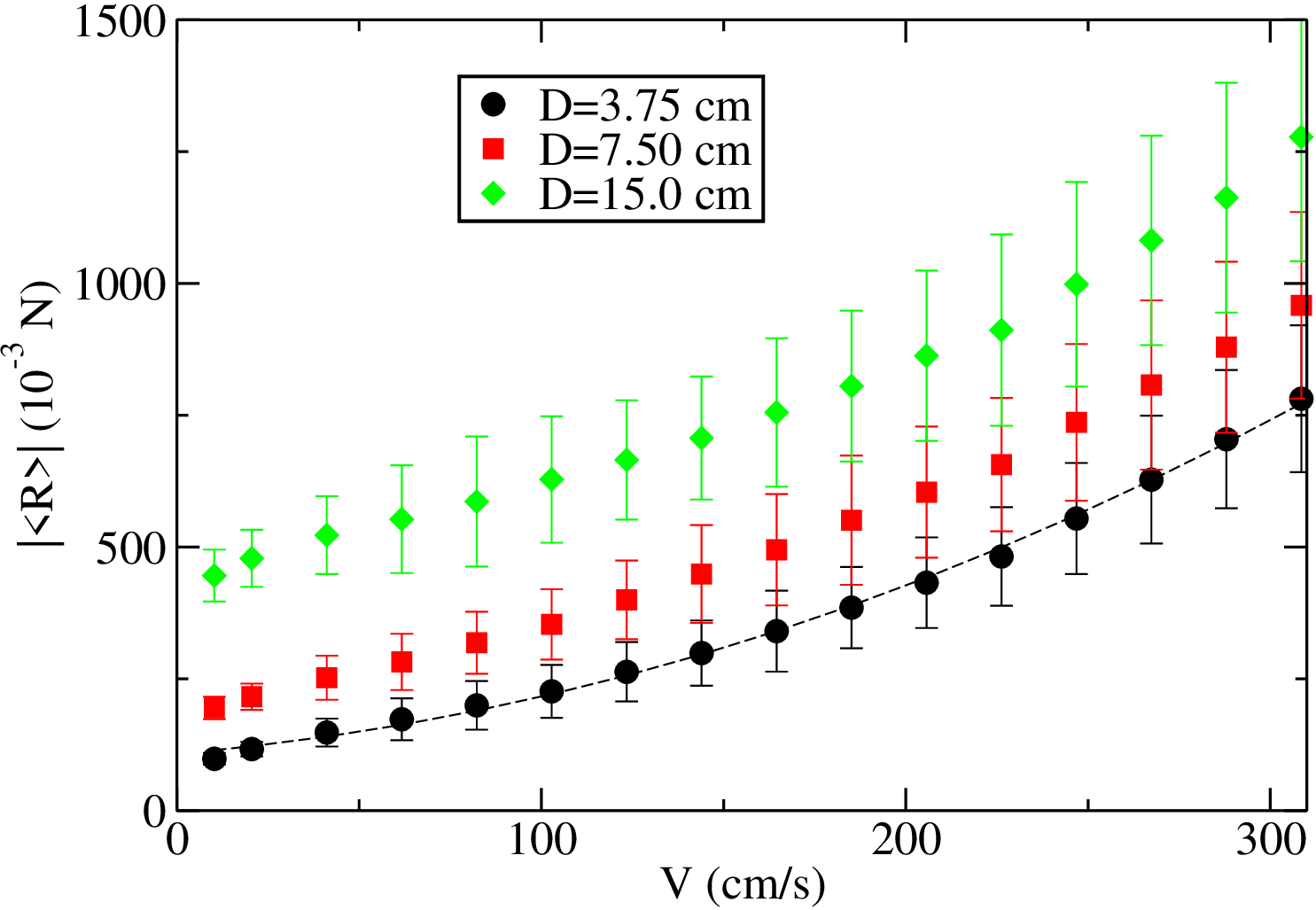,width=13.0cm,height=13.0cm}}
\caption{Drag force on the circle as a function of the drag speed and for 
three distinct depths. Dashed line is a quadratic fit to the $D=3.75$ cm data.
\label{drag_by_V_circle}}
\end{figure}

\begin{figure}[h]
\rotatebox{0}{\epsfig{file=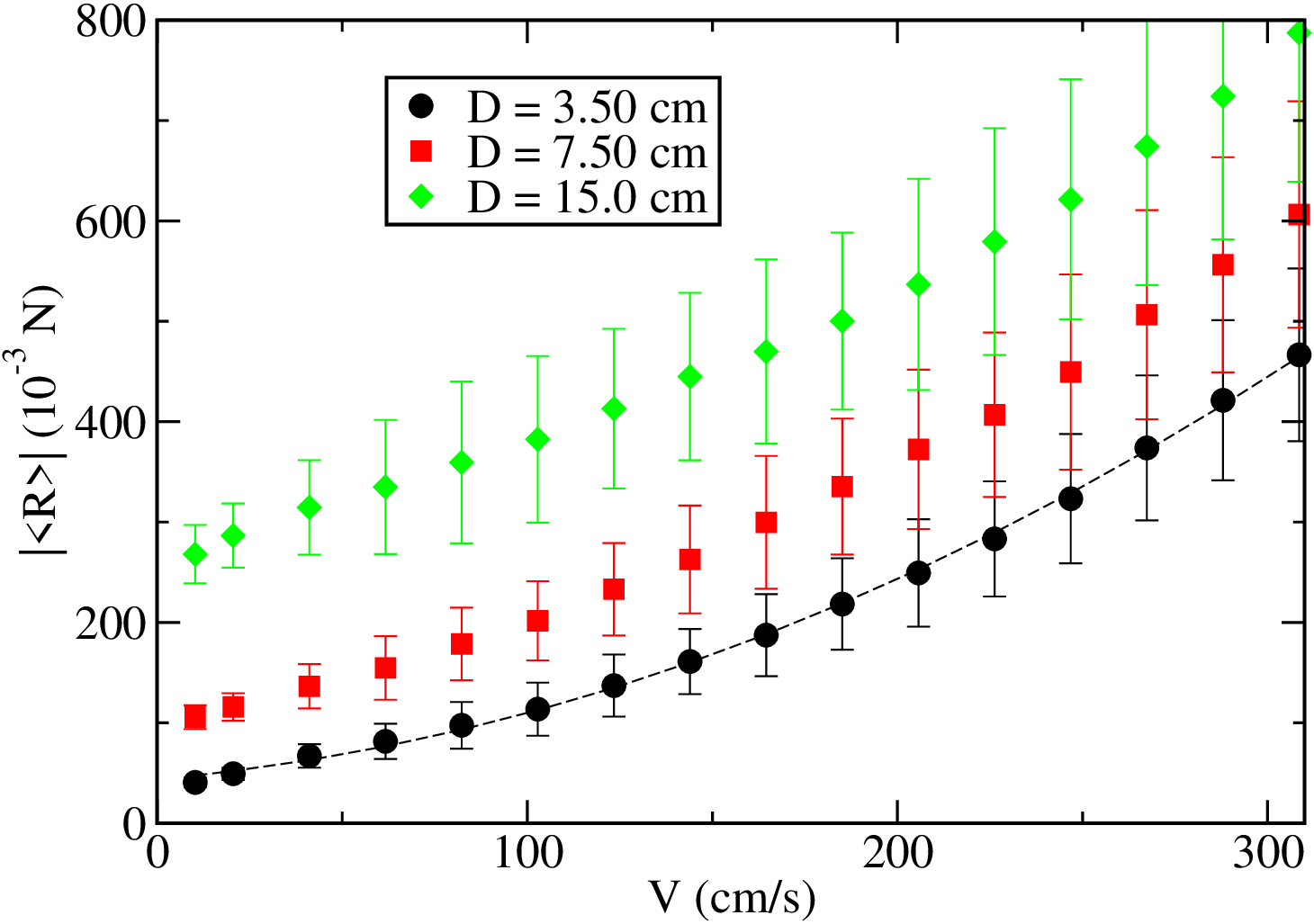,width=13.0cm,height=13.0cm}}
\caption{Drag force on the half-circle as a function of the drag speed and 
for three distinct depths. Dashed line is a quadratic fit to the $D=3.75$ cm 
data.
\label{drag_by_V_half}}
\end{figure}

\begin{figure}[h]
\rotatebox{0}{\epsfig{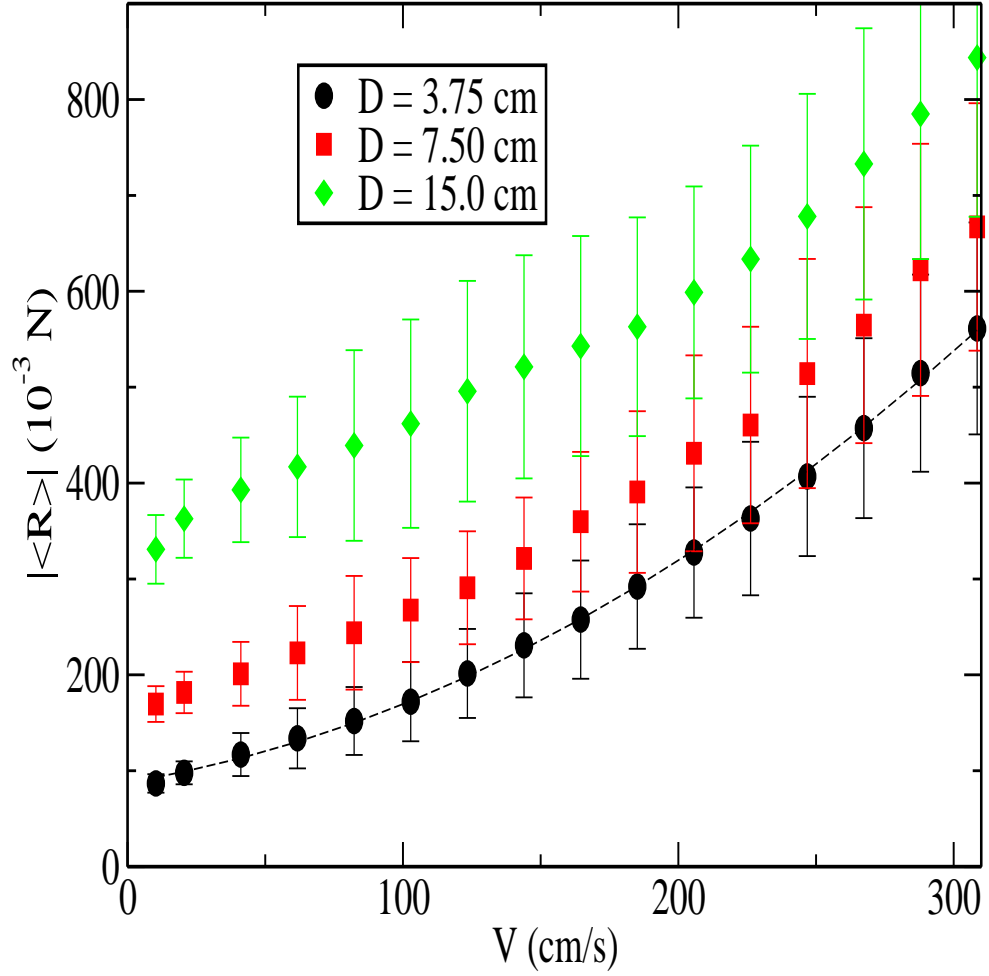}}
\caption{Drag force on the inverted half-circle as a function of the drag 
speed and for three distinct depths. Dashed line is a quadratic fit to the $D=3.75$ cm data.
\label{drag_by_V_half_inverted}}
\end{figure}

We begin with the results for the drag as a function of the drag speed, which 
are shown in figs. 
\ref{drag_by_V_circle} (circle), \ref{drag_by_V_half} (half-circle), 
\ref{drag_by_V_half_inverted} (inverted half-circle). We see clearly that the 
drag monotonically increases 
with $V$. Also, this growth is approximately quadratic: the dashed 
lines in all three figures are quadratic fits to the $D=3.75$ cm curves. Fits 
to the other curves were made as well and show similar agreement. Therefore, 
this relationship is barely affected by shape and depth of the 
intruders. The magnitude of the forces, however, are largest for the circle, 
while they are approximately equal for both half-circles (with the forces on 
the inverted one slightly larger). The quadratic dependence on $V$ indicate 
that the drag follows a similar law to the one of the drag on a sphere dragged 
through a viscous fluid at high Reynolds number \cite{Buchholtz98,Wass03,landau}, i.e., the 
average net drag comes, mainly, from inertial effects (momentum exchange). In 
all three figs., the 
results for $D=15.0$ cm, especially at low speeds, show evident deviation from 
the quadratic $V$ dependence. We will show below that the inertial regime is 
characterized by flow detachment, which means that when the intruder is 
dragged through the packing, there is a trail of very low density left behind 
it. Clearly, this trail will appear at a speed that is dependent on the amount 
of material above the intruder. Therefore, we may understand these deviations 
as being due to the larger pressure at larger depths which makes the inertial 
regime to set in at higher speeds than those at more shallow depths. Finally, 
more studies are needed to fully characterize the transition to the inertial 
regime, a task that is postponed to future work. 

We saw that the average drag bears a quadratic relationship to the drag speed, 
and that it is similar to the behavior of the drag force on immersed bodies in 
fast, viscous flows with the flow speed. The classical fluid dynamics result 
also linearly relates the drag force to the size of the intruder. To establish 
that this relationship is the granular analog of the drag law, we 
should investigate the drag dependence on the intruder size to check for 
linearity between these two quantities. Preliminary results show that, indeed, 
the drag is linear in $d_I$.

\begin{figure}[h]
\rotatebox{0}{\epsfig{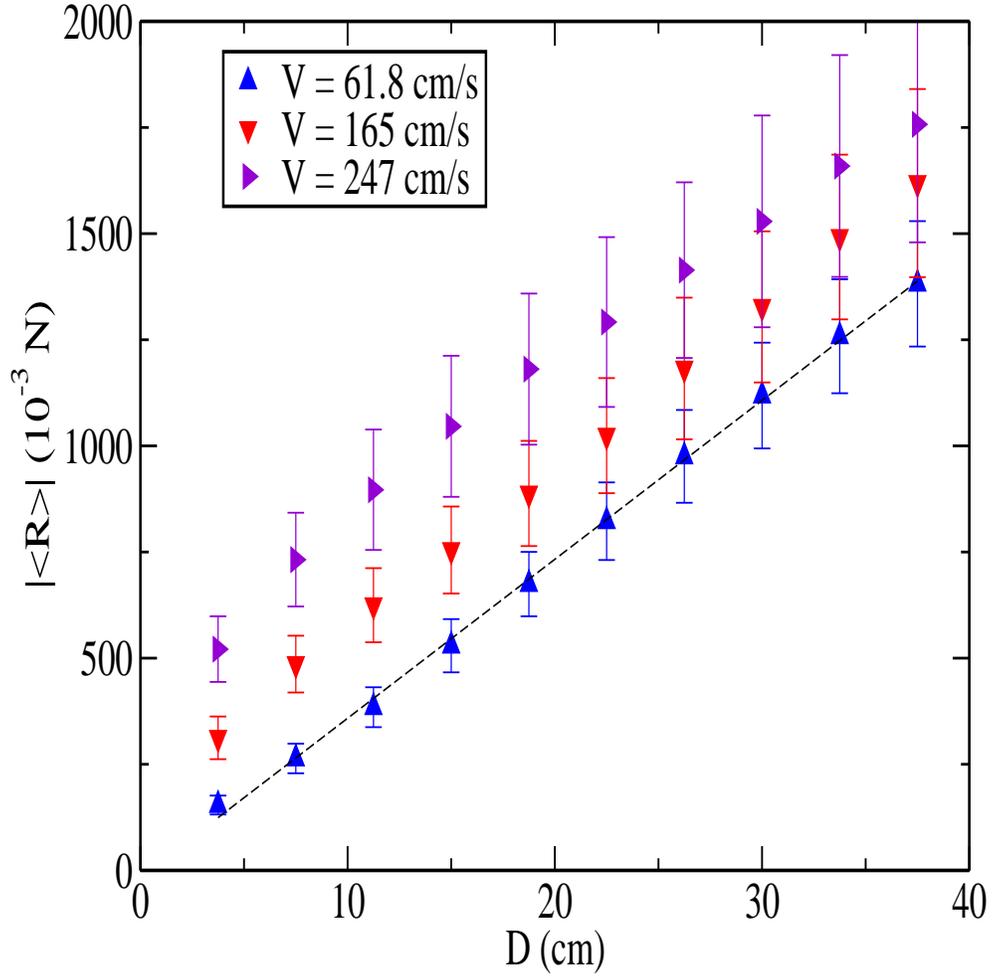}}
\caption{Drag on the circle for varying depths and three distinct drag speed values. Line is a linear fit to the data.
\label{drag_depth_circle_1}}
\end{figure}

\begin{figure}[h]
\rotatebox{0}{\epsfig{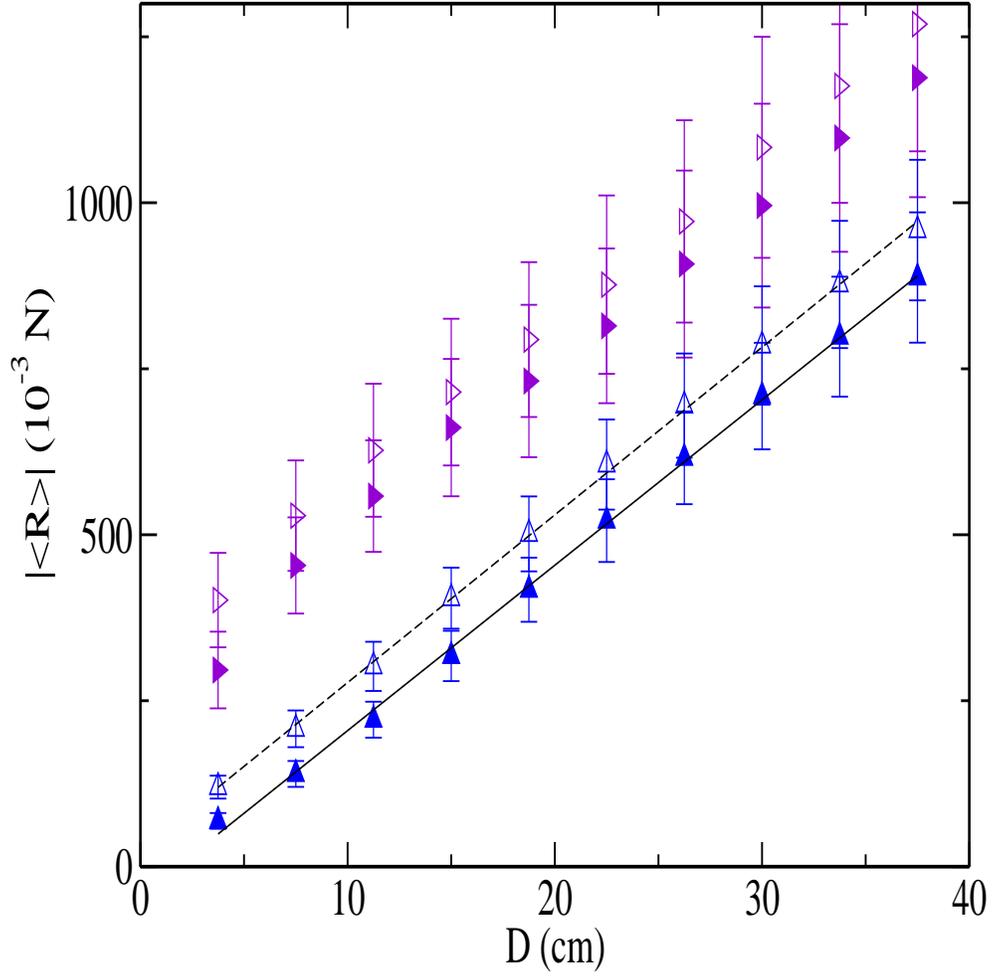}}
\caption{Drag on the half-circle (solid symbols) and inverted half-circle(open 
symbols) for varying depths and two drag speeds (in cm/s): $61.7$ (triangles) 
and $247$ (right triangles). Lines are linear fits to the data.
\label{drag_depth_both_1}}
\end{figure}

The dependence of drag on depth is shown in figs. \ref{drag_depth_circle_1} 
(circle), \ref{drag_depth_both_1} (both half-circles). Like the results for 
$V$, the curves are qualitatively insensitive to shape. We see that the plots 
follow a linear dependence on $D$. The three linear fits, one for each 
shape, confirm this fact. This linear $D$ dependence can be explained by the 
hydrostatic pressure, typical for our packing conditions. Finally, we see, 
again, that the drag force for the circle is larger compared to those for 
both half-circles, which are close to each other. We understand this as a 
consequence 
of the circle's larger projected area perpendicular to the flow, which allows 
it to push more grains than the other two shapes. 
We cannot, however, take the net force on the circle as the 
sum of the forces on the half-circles because of large forces at the leading 
edge of these shapes, a fact seen previously \cite{Ding10} and in our 
force profiles. We can only make this connection if, when summing up 
half-circle forces, do not take into account contributions from the flat sides 
and edges.


\subsection{Lift force\label{lift}}
\begin{figure}[h]
\rotatebox{0}{\epsfig{file=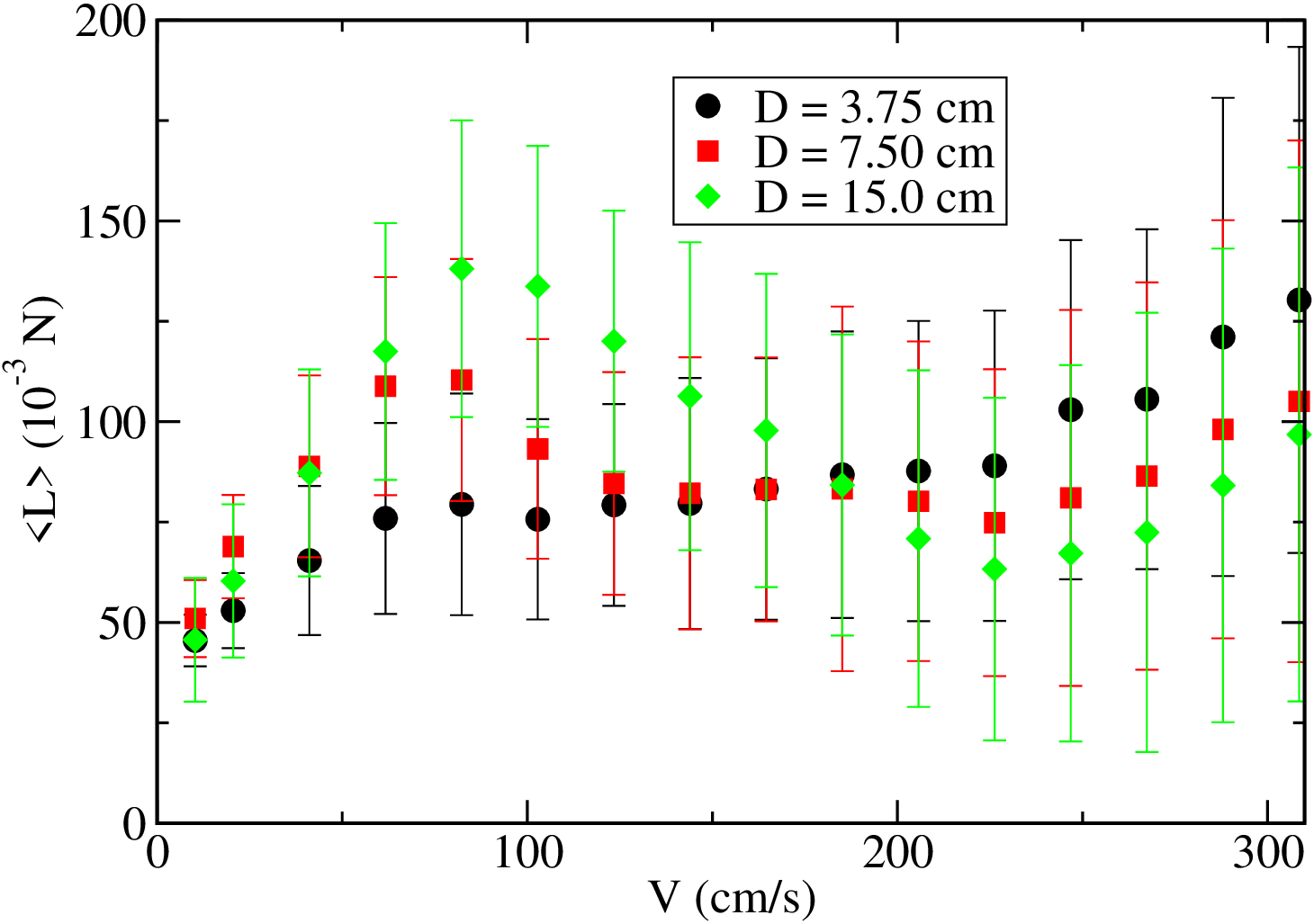,width=13.0cm,height=13.0cm}}
\caption{Lift force on the circle as a function of the drag speed and for 
three distinct depths. 
\label{lift_by_V_circle}}
\end{figure}

\begin{figure}[h]
\rotatebox{0}{\epsfig{file=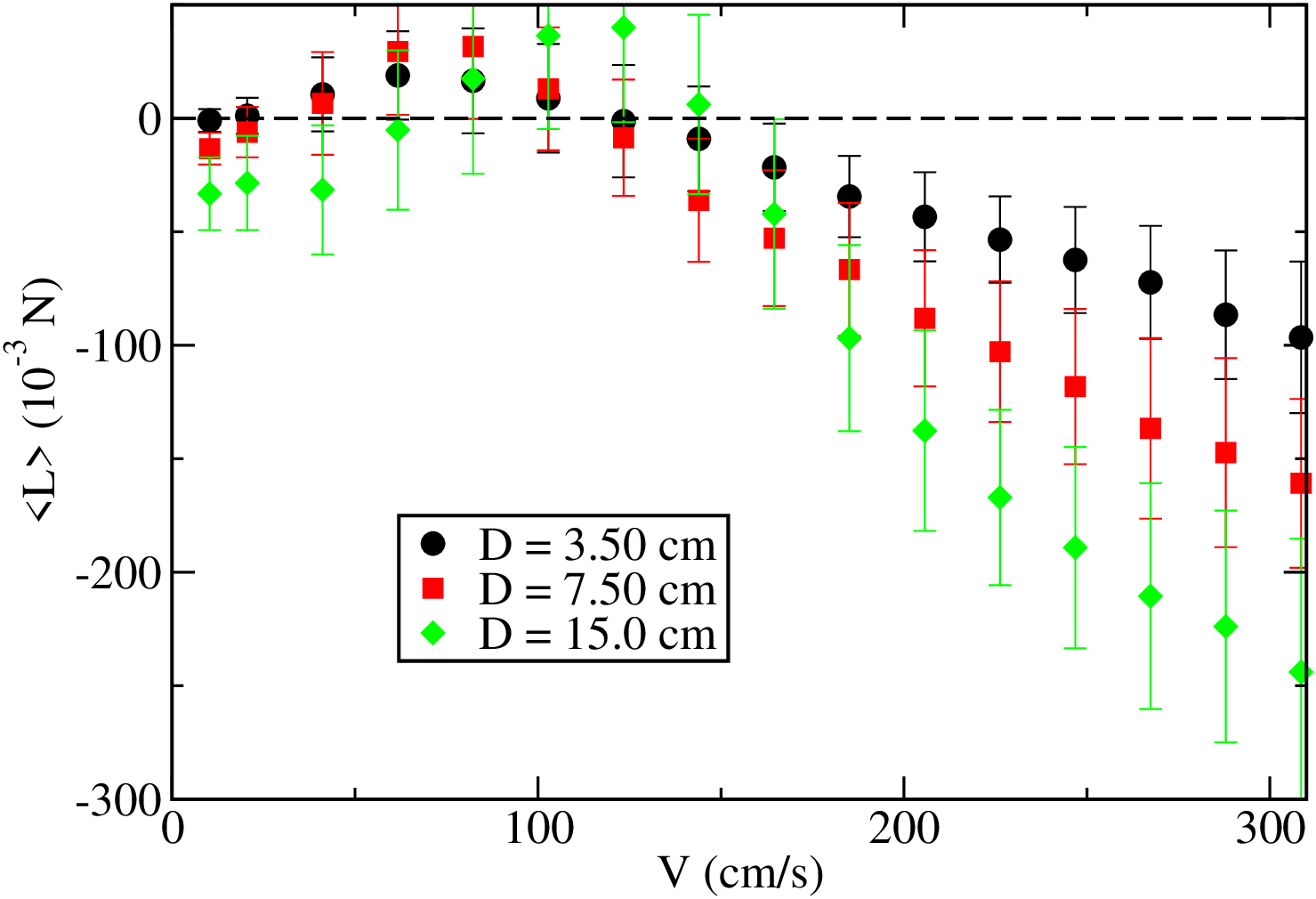,width=13.0cm,height=13.0cm}}
\caption{Lift force on the half-circle as a function of the drag speed and 
for three distinct depths. 
\label{lift_by_V_half}}
\end{figure}

\begin{figure}[h]
\rotatebox{0}{\epsfig{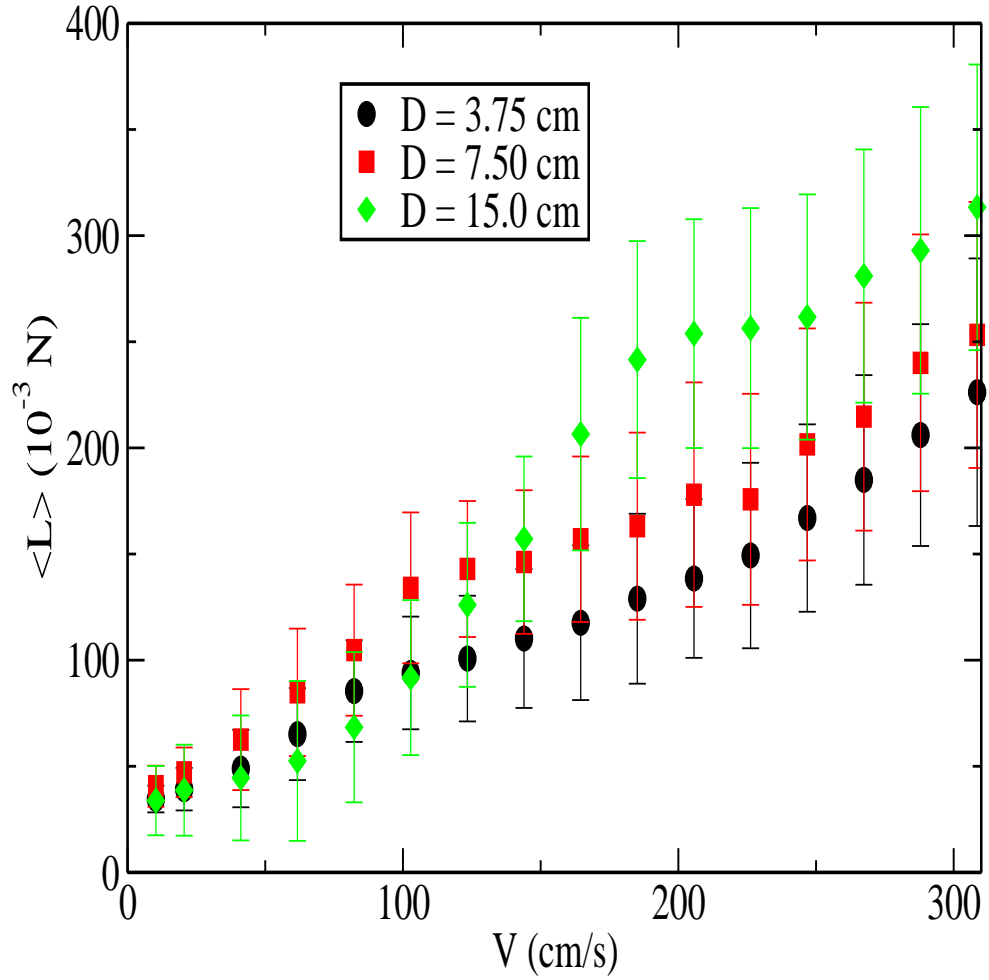}}
\caption{Lift force on the inverted half-circle as a function of the drag 
speed and for three distinct depths. 
\label{lift_by_V_half_inverted}}
\end{figure}

The lift results are shown in figs. \ref{lift_by_V_circle} (circle), 
\ref{lift_by_V_half} (half-circle) and \ref{lift_by_V_half_inverted} (inverted 
half-circle) as functions of the drag speed.

These plots are in marked contrast to those of the drag forces, figs. 
\ref{drag_by_V_circle}, \ref{drag_by_V_half}, 
\ref{drag_by_V_half_inverted}. As stated in the introduction, the behavior of 
the lift with drag speed is complex, so let us begin with the circle. 

We can see that the curves have three distinct regimes: one at low speeds, in 
which $\left<L\right>$ grows with $V$. The range of this regime seems to be 
longer for deep intruders: the curve for $D=15.0$ cm change behavior at 
$V=82.4$ cm/s, while the other two change at $V=61.8$ cm/s; a second regime at 
intermediate speeds, where the lift may decrease (deep intruder) or be roughly 
constant (shallow intruders); and a third regime, appearing at high speeds, 
$V\geq227$ cm/s for all depths, shows $\left<L\right>$ monotonically 
increasing with $V$, and this growth seems to be linear (we need more data to 
confirm that this relationship is linear).


For the half-circle, at low drag speed the lift grows with the $V$, and, 
interestingly, this growth is such that the 
lift force inverts its direction at some depth dependent speed. After 
$\left<L\right>$ reaches a maximum (positive) value, it enters the second 
regime, where it decreases with $V$ and becomes negative again.

For the inverted half-circle, fig. \ref{lift_by_V_half_inverted}, the lift 
grows with the drag speed for all depths and speeds. In any case, we still 
have an initial regime of monotonic growth with $V$, an intermediate regime 
(not very clear for the shallowest intruder), and a fast regime, in which the 
curves are similar to those of the circle.

\begin{figure}[h]
\rotatebox{0}{\epsfig{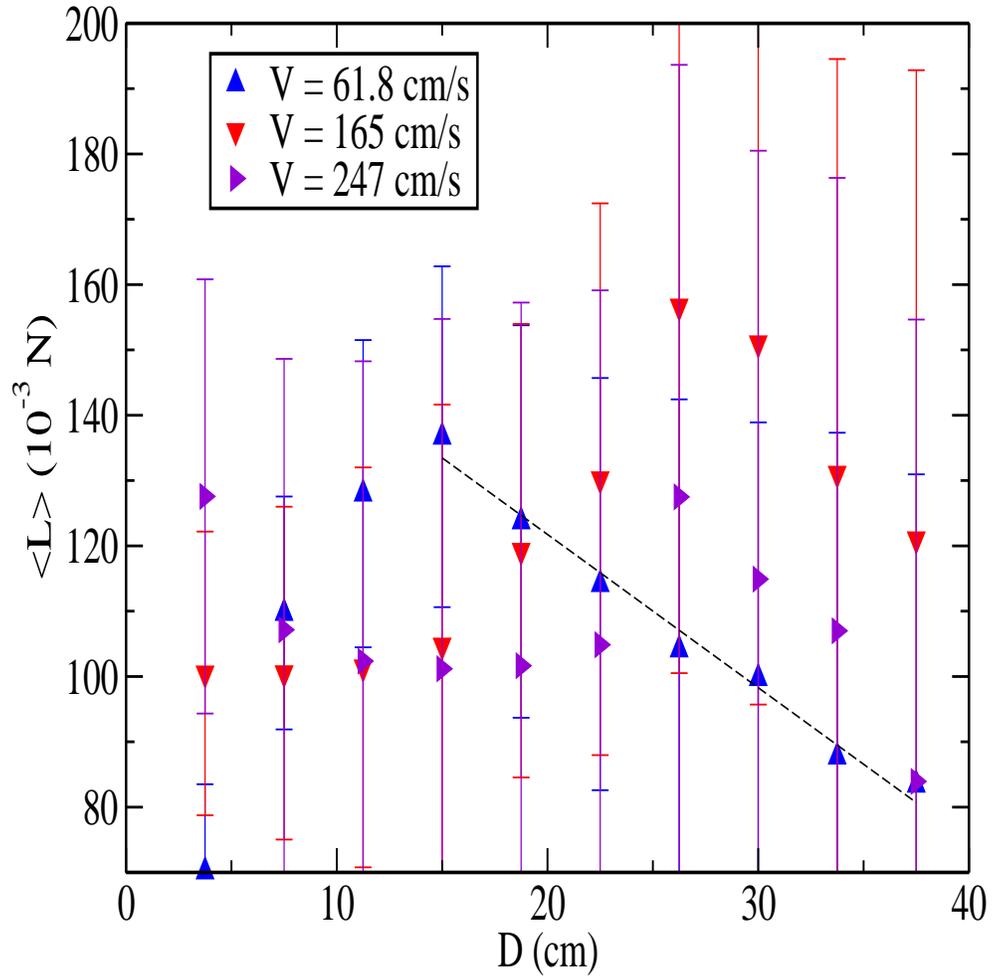}}
\caption{Lift on the circle as a function of intruder depth. 
\label{lift_depth_circle_1}}
\end{figure}

The dependence of $\left<L\right>$ on depth for the circle is shown in fig. 
\ref{lift_depth_circle_1}. We see that, in all three speeds, the lift linearly 
decreases for high $D$, as seen from the linear fit to the $V=247$ cm/s curve. 
For the slowest intruder, this depth is $D=15.0$ cm while for the other two, 
$D=26.3$ cm. The other part of these curves show that, as we increase $V$, 
the lift passes from an initial increase, at $V=61.8$ cm/s, to a plateau, at 
$V=165$ cm/s, to a initial decrease, at $V=247$ cm/s, with $D$. The last two, 
however, will eventually grow and reach a maximum before entering the 
decreasing regime.

\begin{figure}[h]
\rotatebox{0}{\epsfig{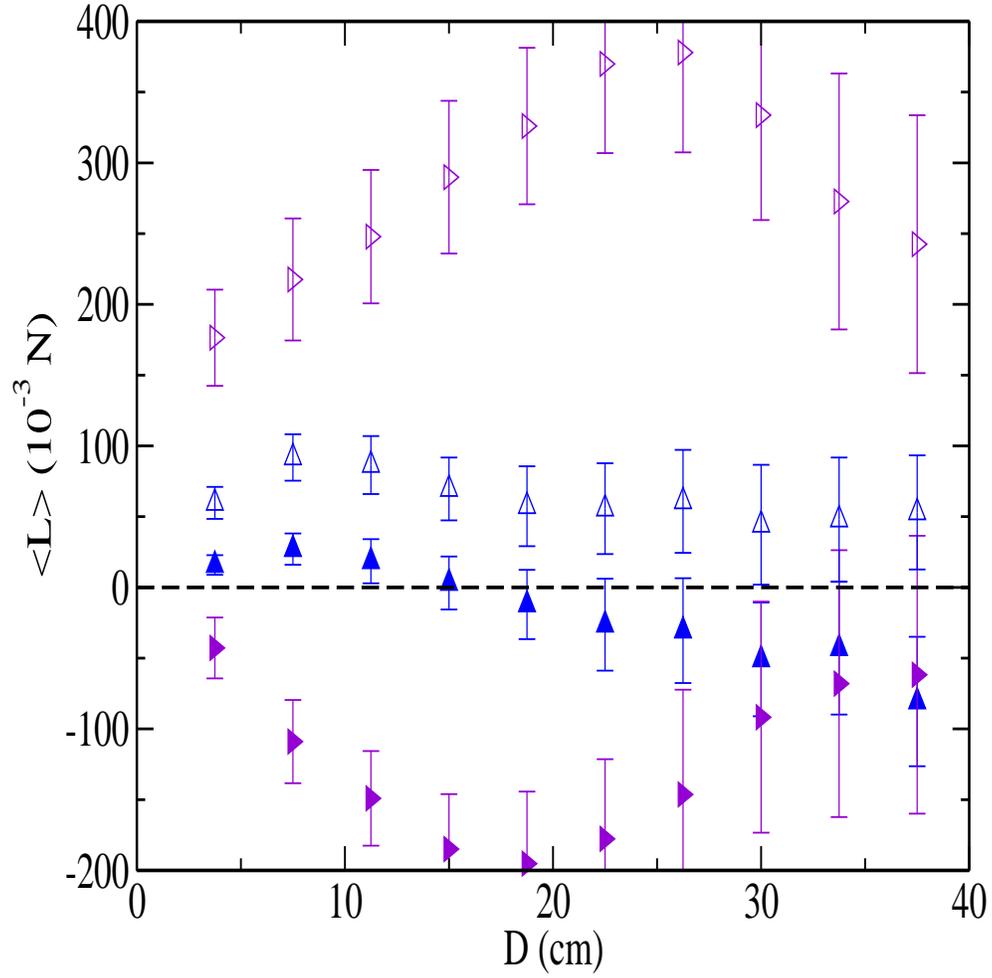}}
\caption{Lift on the half-circle (solid symbols) and inverted half-circle (open 
symbols) for varying depths and two drag speeds: $61.8$ cm/s (triangles) and 
$247$ cm/s (right triangles). 
\label{lift_depth_both_1}}
\end{figure}

The curves for both half-circles, fig. \ref{lift_depth_both_1}, are markedly 
different from those of the circle. First, we see that only for inverted 
half-circle at high speeds the 
lift behaves qualitatively similar to the one in the circle. For a fast 
half-circle, the curve has the inverse relation, it decreases at low depth up 
to minimum at $D=18.8$ cm, and increases for larger $D$. For $V=61.8$ cm/s, the 
lift on the inverted half-circle is not much affected by the depth, except at 
low $D$. Finally, we see again the sign inversion phenomenon seen in fig. 
\ref{lift_by_V_half}: the lift is positive, although small, 
for depths below $D=15$ cm, and negative above this value.

All these results show a very interesting picture: the drag has simple 
relations with speed and depth, while the lift does not. Both of these forces, 
however, come from the same kind of interactions, which are grain-intruder 
repulsive 
contacts. Therefore, we naturally ask the reasons behind these distinct 
relations. We know that the forces depend on the number of contacts and the 
overlap in these contacts (friction is also proportional to the 
overlap, but is bounded due to Coulomb failure condition, so its influence is 
limited). These two quantities 
are not, necessarily, related: a large force could result from a large number 
of contacts with small overlaps or from a few contacts with strong overlaps. 
Moreover, a 
knowledge of both quantities, number of contacts and contact overlaps, does 
not explain our results. We should also pay 
attention to the contact position in the intruder surface and to the fact that 
we measure net forces. The first information is 
needed because, as shown in \cite{Ding10}, the forces depend on the local 
geometry. The second is important because the drag is the sum of the 
horizontal forces on the leading and back sides, while the lift is the result 
of contributions from the vertical forces above and below the horizontal line 
(the flow direction) that passes through the center of the intruder. Since the 
results for the net lift have more interesting behaviors, we concentrate our 
efforts in understanding the reasons that led to the results of figs. 
\ref{lift_by_V_circle} to \ref{lift_depth_both_1}. Hence, the following 
analysis deals mainly with data for lift force.

We reasonably assume that the contact overlaps on the leading side increase 
with $V$, and those on the back side decrease with speed. Also, overlaps should 
increase over the whole intruder for larger depths due to larger packing 
pressure. Since the drag always increases with $V$ and $D$, we can easily 
understand the drag results as the leading drag increasing faster with 
increasing speed and depth than that on the back side for the whole range of 
parameters we studied. 

Following this explanation, we conclude that the lift forces above and 
below the intruder are comparable, since the dependence on $V$ and $D$ is 
non-monotonic. The decreasing regimes seen in figs. 
\ref{lift_by_V_circle}, 
\ref{lift_by_V_half}, \ref{lift_depth_circle_1} and \ref{lift_depth_both_1} 
can only occur if the lift on top increases faster with $V$ and $D$ 
than that on the bottom. In the same way, increasing lift regimes only occur 
if vertical forces on top grow faster with $V$ and $D$ than those on the 
bottom. Since the overlaps on the bottom are larger than those on the 
top, the only way to have decreasing regimes 
is if some effect occurs with the contact number. Moreover, this effect is 
not, in principle, the same 
above and below the intruder. We need to understand the effects of changing 
speed and depth on the contact number in order to fully understand our results. 
This is done in the next section where we present data on the lift and contact 
profiles as well as their contribution from each side of the intruder.

\section{Fields and profiles\label{fields}}
From the previous discussion, the nonmonotonic relationship between 
$\left<L\right>$ and $V$ is the consequence of the unequal, but comparable, 
forces on the upper and lower sides of an intruder. Also, these two 
contribution should have their own relationships with $V$ and $D$. To have a 
more clear idea of the effects into play, we show in this section data on the 
lift and contact profiles along the intruder surface, as well as velocity 
fields.

\begin{figure}[h]
\rotatebox{0}{\epsfig{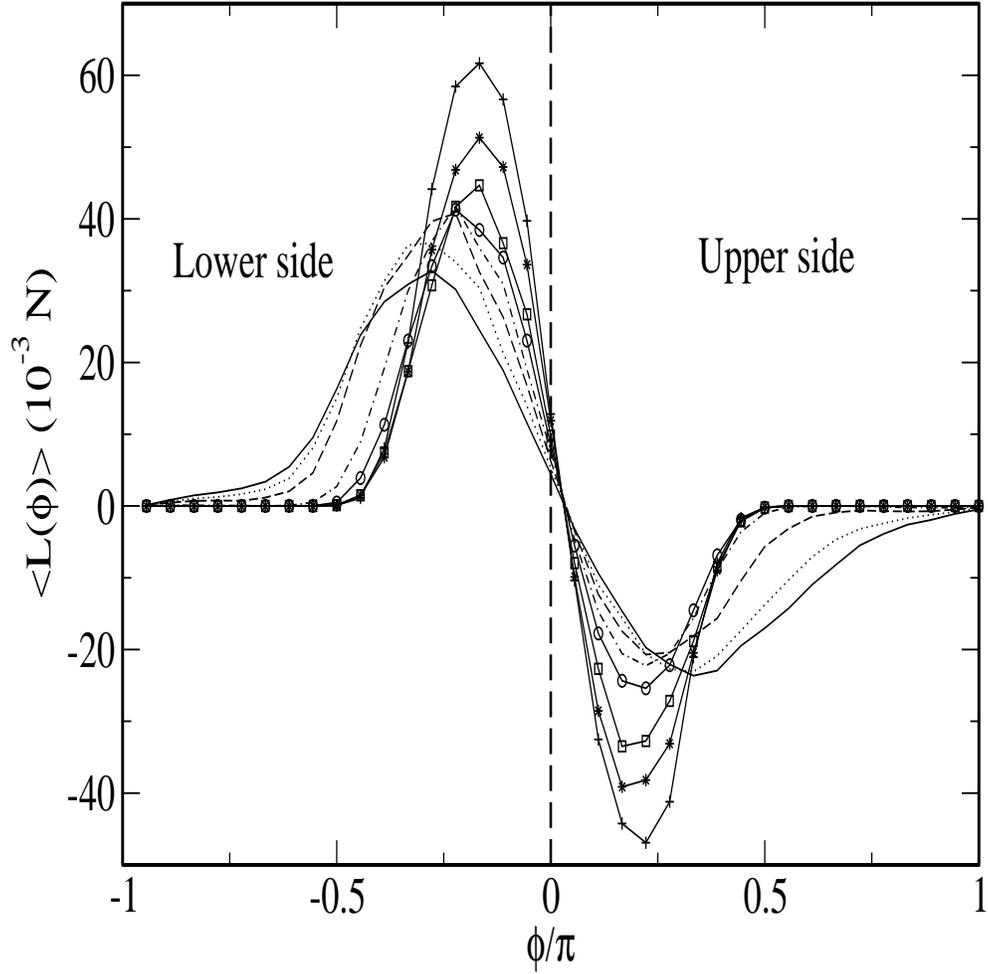}}
\caption{Lift profiles on the circle for $D=15.0$ cm. Speeds (in cm/s): 
$10.2$ (solid), $41.2$ (dotted), $82.4$ (dashed), $124$ (dot dashed), $165$ 
(circles), $227$ (squares), $268$ (stars), $309$ (crosses).
\label{prof_lift_circle_1}}
\end{figure}

Let us begin with the lift profiles, $\left<L(\phi)\right>$. A few examples 
are shown in fig. \ref{prof_lift_circle_1} for the circular intruder 
for $D=15.0$ cm and $8$ distinct speed values. We first see that these 
profiles are consistent with previous numerical results \cite{Ding10}, which 
showed 
that the force is larger on the lower side than in the upper side (hydrostatic 
condition) and it is not symmetric with respect to the leading point in the 
intruder, which allows a vertical plate to have a nonvanishing, positive lift 
when dragged through the packing.

\begin{figure}[h]
\rotatebox{0}{\epsfig{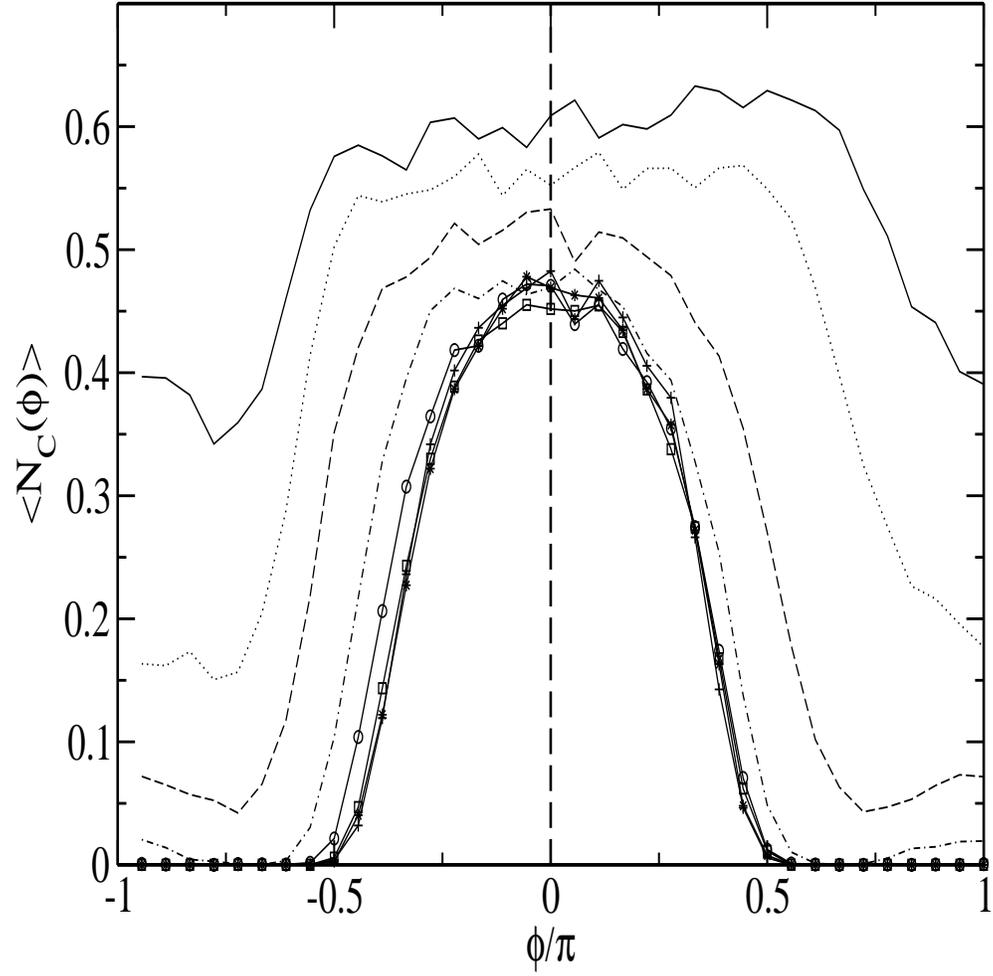}}
\caption{Contact number profiles for the circle for $D=15.0$ cm. 
Speeds (in cm/s): 
$10.2$ (solid), $41.2$ (dotted), $82.4$ (dashed), $124$ (dot dashed), $165$ 
(circles), $227$ (squares), $268$ (stars), $309$ (crosses).
\label{prof_cont_circle_1}}
\end{figure}

\begin{figure}[h]
\rotatebox{0}{\epsfig{file=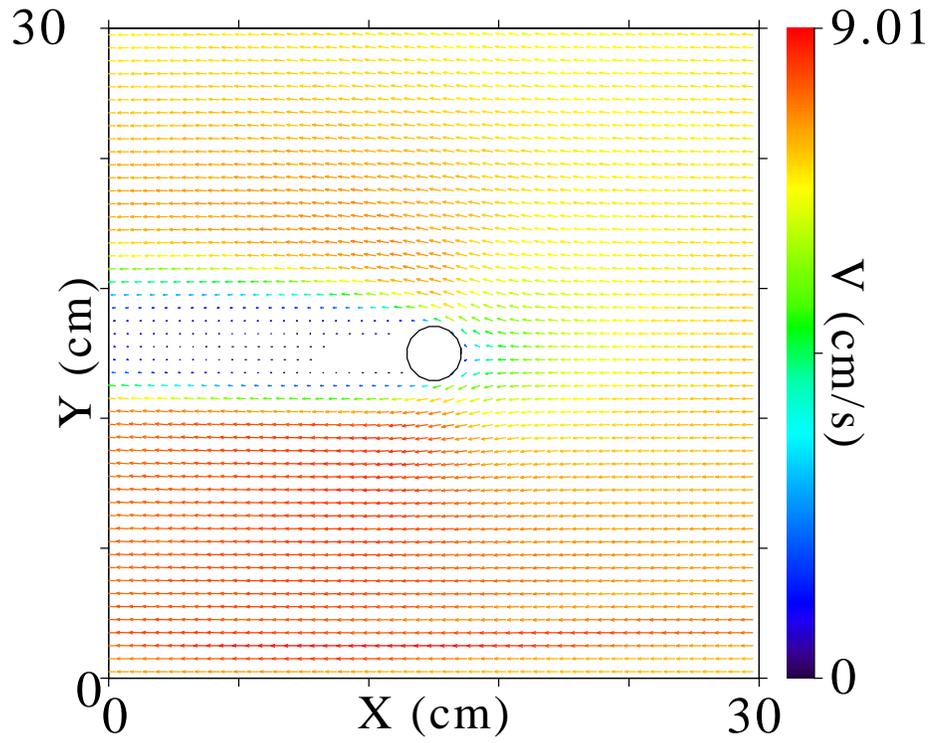,width=13.0cm,height=13.0cm}}
\caption{Velocity field, measured in the intruder's frame, in a window $30.0$ cm 
$\times$ $30.0$ cm centered on the circle, at $V=268$ cm/s and $D=15.0$ cm.
\label{vel_field_circle_1}}
\end{figure}

\begin{figure}[h]
\rotatebox{0}{\epsfig{file=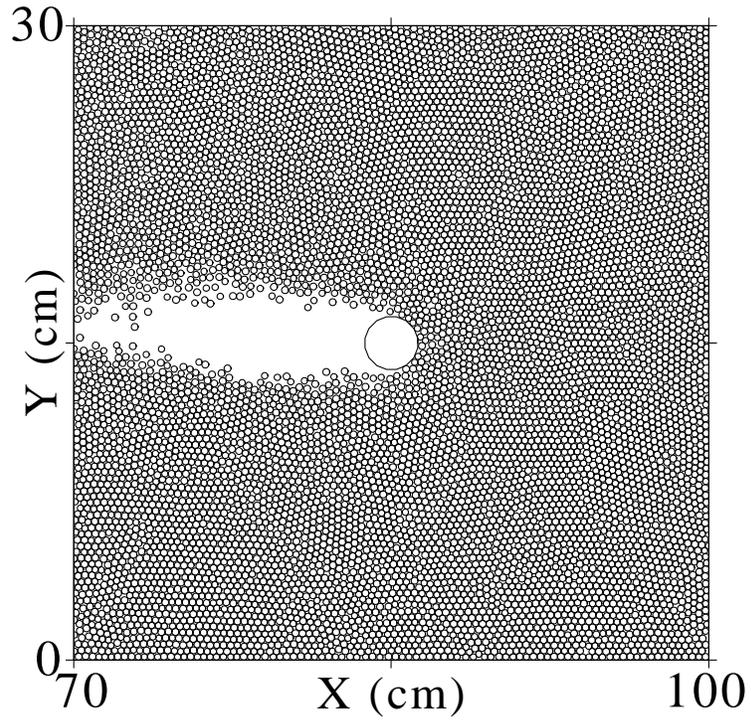,width=11.0cm,height=11.0cm}}
\caption{Local packing configuration around the circle at $V=247$ cm/s and $D=15.0$ cm.
\label{config_circle_1}}
\end{figure}

Another interesting aspect of these profiles is that, as we increase the 
drag speed, the forces become more concentrated in the range 
$[-\pi/2,\pi/2]$, with practically zero force outside this range. This features 
is also seen in the drag profiles (not shown), $\left<R(\phi)\right>$, which 
are symmetrical with respect to $\phi=0$ and are peaked around this value (the leading end of the intruder). However, in this 
case, the non-zero elements of the profiles have all the same sign (negative) 
instead of the positive and negative contributions seen in fig. \ref{prof_lift_circle_1}. This fact naturally leads to a larger net drag compared to a net 
lift. We conclude, then, that, at high speeds, the 
net drag and lift come from contribution made only on the leading side of the 
intruder. At low speeds, the forces are nonzero over the whole surface (a zero 
force value 
in the profile means that no grain ever touched the intruder at that point). To 
confirm this fact, we show in fig. \ref{prof_cont_circle_1} the contact number 
profiles for the same cases as those of fig.  \ref{prof_lift_circle_1}. 
Indeed, we see that our previous statement on the concentration of forces on the leading side of the intruder is confirmed. We show, in figs. 
\ref{vel_field_circle_1} and \ref{config_circle_1}, examples 
of velocity field and packing configuration for high speed. We can 
see that the zero contact number is a consequence of to the appearance of a 
trail of zero density behind the intruder. Since when it moves it pushes 
grains up and down, the trail is formed because these thrown grains do not 
fall back to their initial heights fast enough to fill up the space behind 
the intruder. It is easy to see that at lower drag 
speeds, grains will be pushed more gently, up to a point where they merely 
pass over the intruder (first regime) and the trail is absent. We observed 
that this effect is depth independent, the trail is always formed 
for large enough speeds. 

The appearance of a vanishing density region behind an 
intruder is commonplace in studies of supersonic granular flow 
\cite{Buchholtz98,Amar01,Rericha02,Chehata03,Wass03,Amar06,Boudet08,Boudet09,Boudet10}. 
Therefore, we call this effect a granular flow analog of detachment of 
supersonic fluid flow around obstacles: at high drag speeds, there are only 
contacts on the leading side of the intruder. The marked distinctions that 
appear in our results concern the shock front. In the cited reports, shock 
fronts are seen to be symmetrical regarding the incident flow direction, with a 
parabolic shape, with a solid region right in front of the leading 
(stagnation) point in the obstacle. Our shock front is asymmetrical with 
respect to the drag direction, since gravity is perpendicular to this 
direction. In addition, we do not observe any parabolic, static configuration 
in front of the intruder, although there is a typical stagnation point there: 
grains are pushed away by the intruder which prevents any such phenomenon to 
occur, see fig. \ref{vel_field_circle_1}.

From this discussion, it is easy to see that the amount of material above the 
intruder decreases the height that grains attain when thrown up by the 
intruder. 
Therefore, the speed at which the trail appears is affected by depth. We 
may estimate this speed by calculating the time, $t_F$, a thrown grain 
takes to fall a distance $h$ and 
the time the intruder takes to move an effective radius $(d+d_I)/2$ (assuming 
that the falling grain and the intruder are at the same horizontal position). 
A precise estimate can only be 
given if we can calculate the distance $h$. Since this is a complicated 
matter, we need to take into account how the forces are transmitted through 
the packing and the effects of depth in order to calculate $h$ 
more accurately, we assume that this height is, simply, 
one effective radius. Then, the time a grain takes to free fall this distance 
is given by: 
\[
t_F=\sqrt{\frac{(d+d_I)}{2g}}.
\]
The time the intruder moves one effective radius at speed $V$ is:
\[
t_D=\frac{d+d_I}{2V}.
\]
The trail occurs when $t_F\geq t_D$. Then, the onset of this phenomenon is 
given by:
\[
V=\sqrt{\frac{(d+d_I)g}{2}}.
\]
From our data, $d=0.32$ cm (mean particle diameter), $d_I=2.54$ cm and $g=981$ 
cm/s$^2$, the speed value can be calculated to be $V=37.5$ cm/s. This value 
is below the speed at which the first regime ends, between $61.7$ cm/s and 
$81.3$ cm/s.

\begin{figure}[h]
\rotatebox{0}{\epsfig{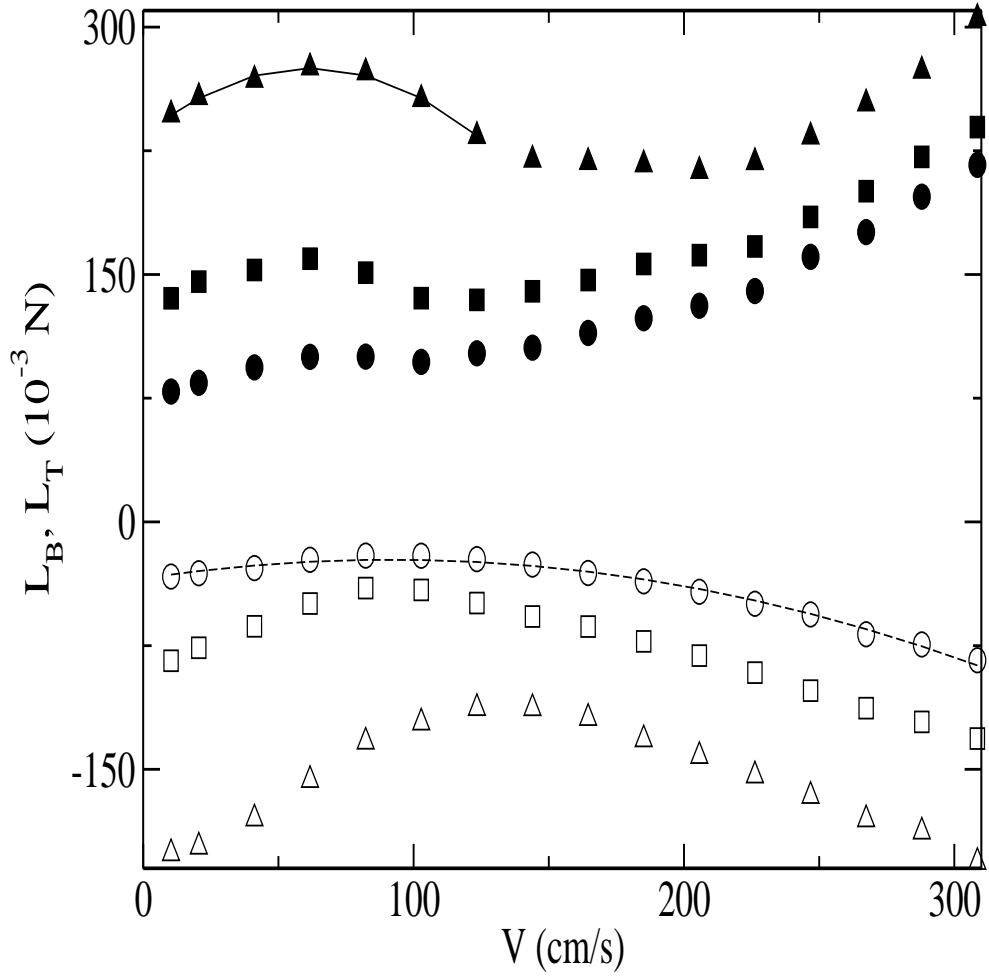}}
\caption{Lift on top (open) and bottom (filled symbols) of the circle as 
functions of the drag speed. Depths (in cm): $3.75$ (circles), $7.50$ 
(squares) and $15.0$ (triangles). Dashed and solid lines are quadratic fits 
to the data.
\label{lift_sides_circle_1}}
\end{figure}

\begin{figure}[h]
\rotatebox{0}{\epsfig{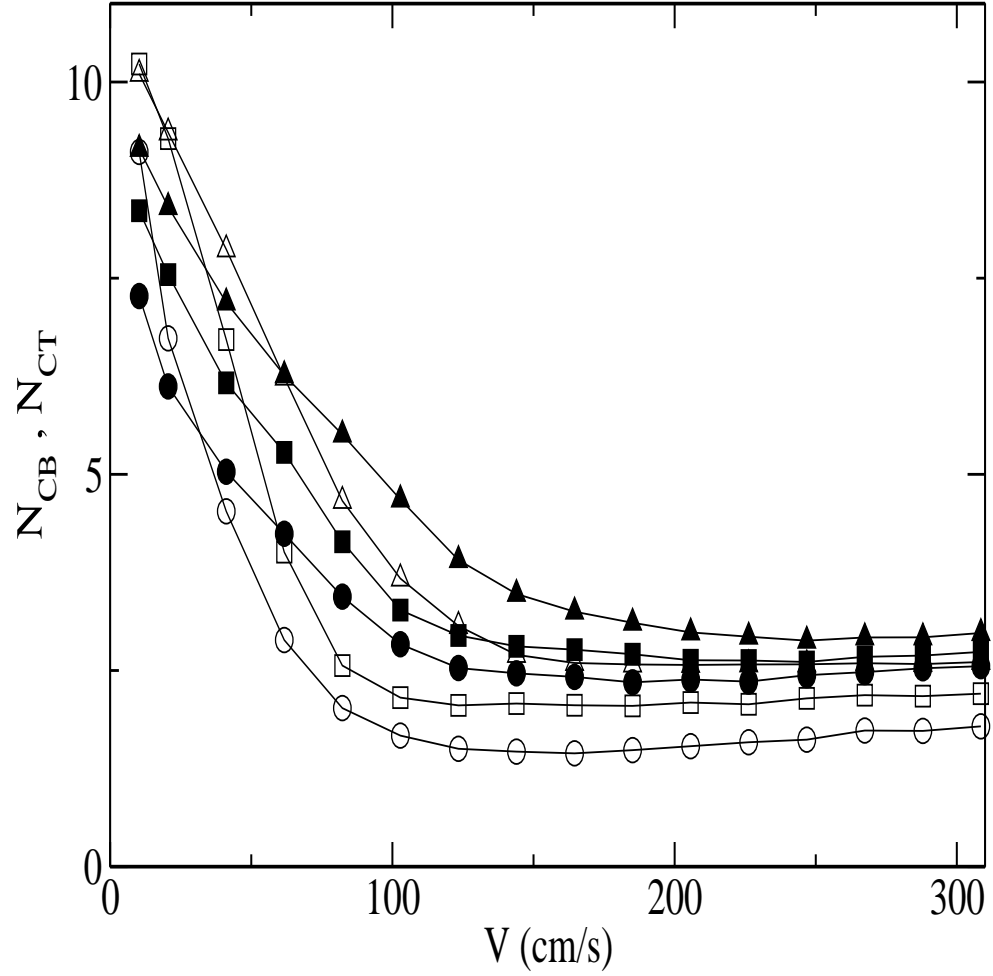}}
\caption{Contact numbers on top (open) and bottom (filled symbols) of the 
circle as functions of the drag speed. Depths (in cm): $3.75$ (circles), 
$7.50$ (squares) and $15.0$ (triangles). Lines are guides to the eyes.
\label{cont_sides_circle_1}}
\end{figure}

In addition, it is clear from fig. \ref{prof_cont_circle_1} that the overall 
contact number decreases with $V$, which confirms our previous argument, the 
number 
of contacts is affected by the increase in drag speed. This change, however, 
does not affect the drag force, since most of its value comes from the leading 
side. The effect is seen on the lift because the forces on the upper and lower 
sides of the circle are comparable. We show in figs. \ref{lift_sides_circle_1} 
and \ref{cont_sides_circle_1} the values of the lift and contact number on top 
and bottom of the circle for all speeds in order to consider each contribution 
separately.

We see that the top lift has an approximate quadratic dependence on $V$, like 
the drag, as suggested by the quadratic fit for the $D=3.75$ cm curve (dashed 
line). The bottom lift, on the other hand, shows only one evident quadratic 
regime at low speeds. This behavior is seen from the quadratic fit to the 
$D=15.0$ cm curve. Given the drag results, we feel that the lift on bottom 
should be quadratic at all $V$ and that, perhaps due to large noise in lift 
measurements, also reported in \cite{Soller06}, such behavior is hidden in 
these plots. In any 
case, we see that the lift behavior at low speeds is correlated to a fast 
decrease of the contact number on both sides. This fast decrease should be 
followed by a drop in the forces, which is seen in top lift. The bottom one, 
however, increases for low $V$. 

\begin{figure}[h]
\centering
\begin{tabular}{cc}
\rotatebox{0}{\epsfig{file=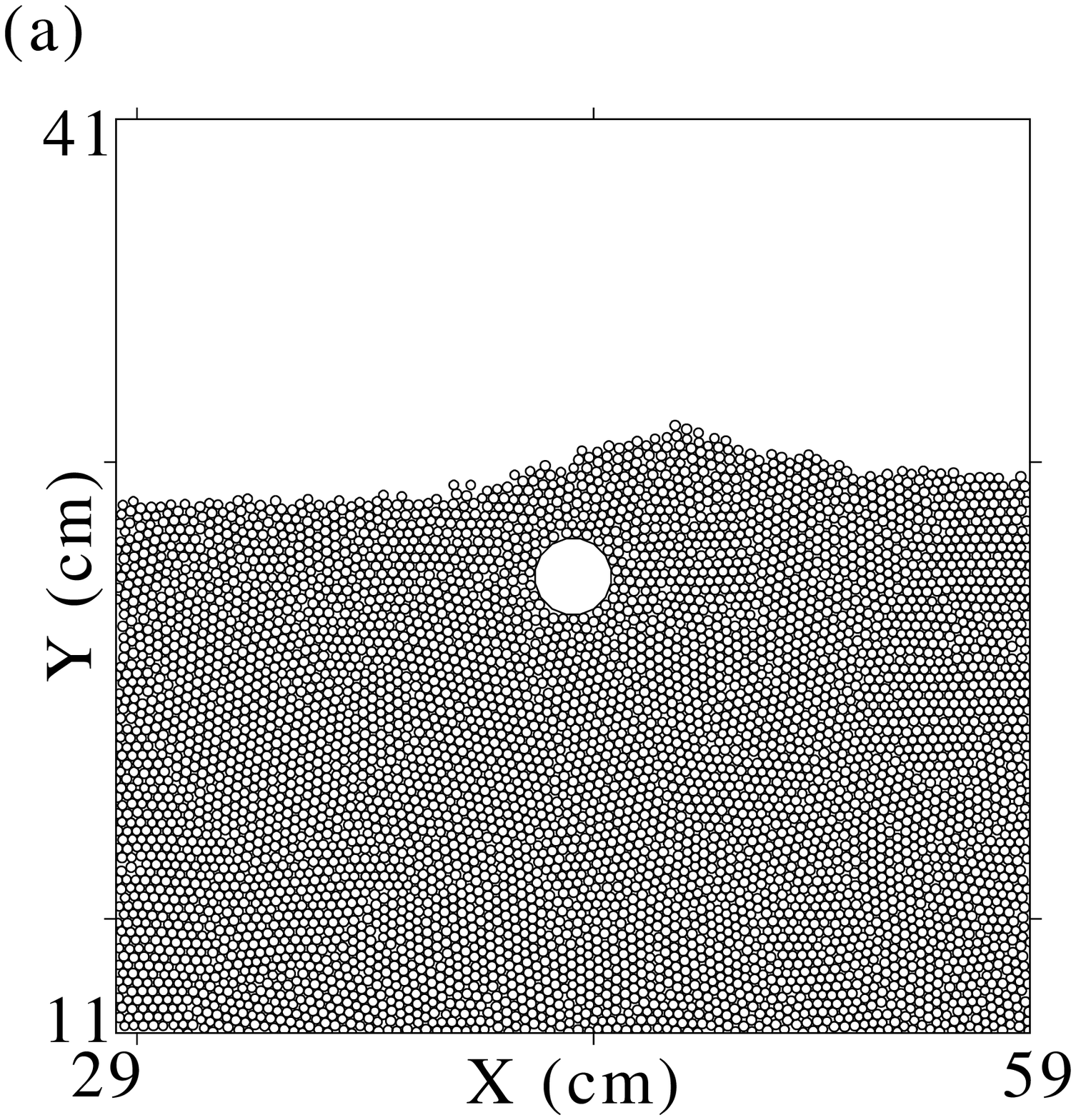,width=7.5cm,height=7.5cm}} & 
\rotatebox{0}{\epsfig{file=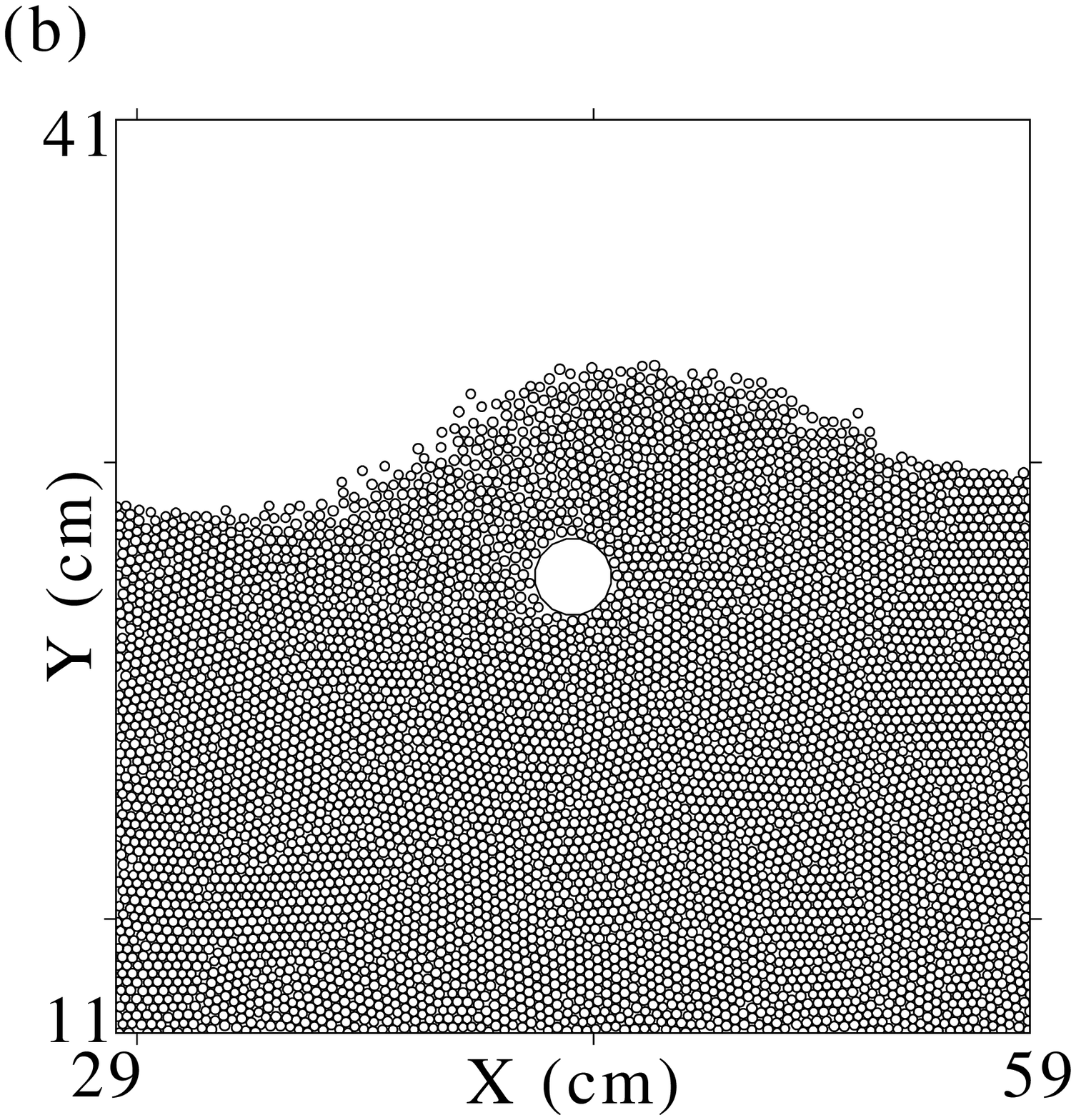,width=7.5cm,height=7.5cm}}\\
\rotatebox{0}{\epsfig{file=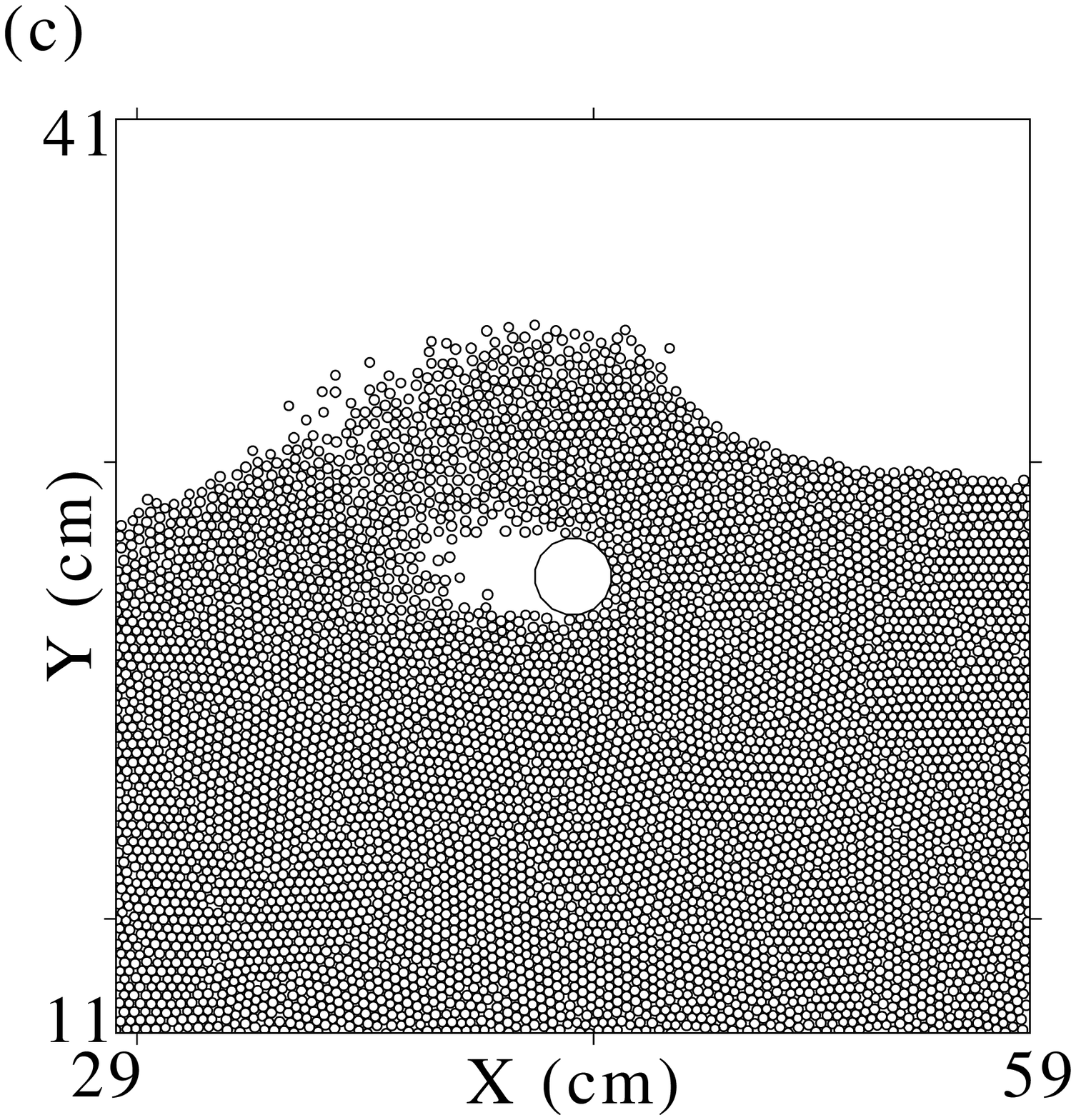,width=7.5cm,height=7.5cm}} & 
\rotatebox{0}{\epsfig{file=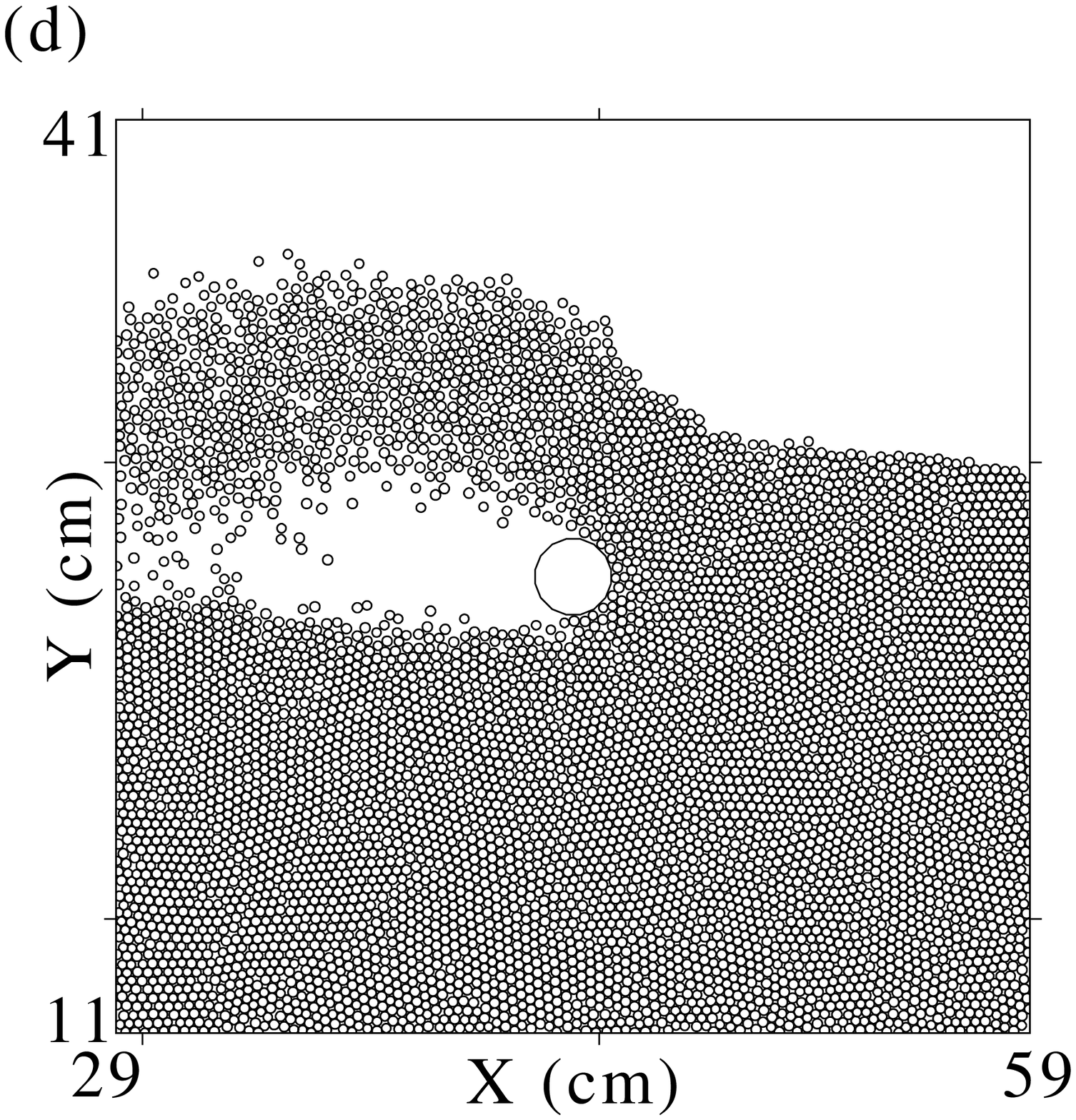,width=7.5cm,height=7.5cm}}\\
\end{tabular}
\caption{Configurations snapshots for $D=3.75$ cm and distinct speeds 
(in cm/s): (a) $10.3$, (b) $61.8$, (c) $103$ and (d) $165$.
\label{configs_middle}}
\end{figure}

This results can be understood by the 
following mechanism, illustrated in fig. \ref{configs_middle}: when the 
intruder moves at low speeds, panel (a), grains flow upward and increase the 
free surface height in front of the intruder \cite{Ding10}. This accumulation 
will increase the packing pressure and, consequently, the downward vertical 
force done by the bottom side. As the drag speed is increased, panel (b), the 
bump in the 
free surface moves from a position in front of the intruder to one right above 
it. This relieves the packing close to the intruder, and the bottom lift 
decreases. As the speed is increased even further, panels (c) and (d), the 
material is thrown up in the air and the trail behind the intruder is formed. 
At high speeds, the contact number reach a plateau, which means that all 
forces depend on $V$ mainly through increasing pressure on the leading side.

\begin{figure}[h]
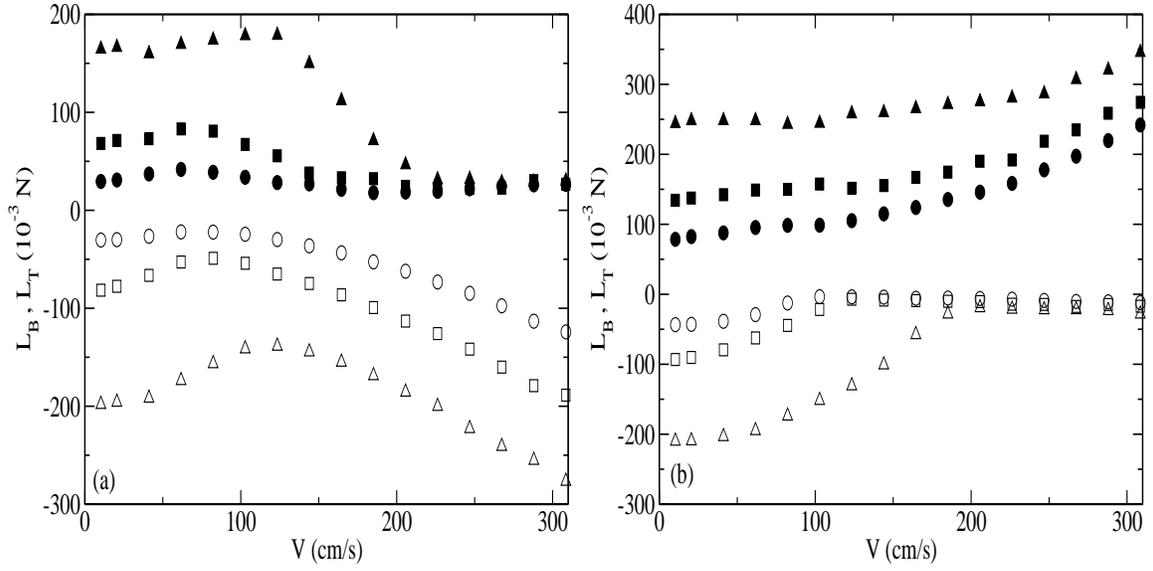

\centering
\begin{tabular}{cc}
\rotatebox{0}{\epsfig{file=Lift_Sides_P7_P8_P9.eps,width=7.5cm,height=7.5cm}} & 
\rotatebox{0}{\epsfig{file=Lift_Sides_P10_P11_P12.eps,width=7.5cm,height=7.5cm}}\\
\end{tabular}
\caption{Lift on top (open) and bottom (filled symbols) of half-circle, panel 
(a), and inverted one, (b), as functions of the drag speed. Depths (in cm): 
$3.75$ (circles), $7.50$ (squares) and $15.0$ (triangles). 
\label{lift_sides_both_1}}
\end{figure}

The conclusions we reached for both sides of the circle also hold for the 
half-circles. Comparing figs. 
\ref{lift_by_V_half} and \ref{lift_by_V_half_inverted} with those for top and 
bottom lift, respectively, we see that they are qualitatively similar. It 
implies that the contacts on the flat side should not have a severe change with 
$V$, even though they are the reason the lift inverts sign for the half-circle. 
The lift and contacts on the flat side show a distinct picture. Our results, 
fig. \ref{lift_sides_both_1}, reveal that the lift force 
on the flat sides, on both half-circle shapes, decrease as we increase the 
speed and reach practically the same 
value at high $V$, independent on depth, although the initial 
values of this force grows with $D$. This high speed plateau is an obvious 
consequence of flow detachment. When the flow begins to detach from the 
intruder, practically all contacts on the flat side will disappear, since there 
is no projected area perpendicular to the flow. Forces will only be exerted on 
the leading point, which barely changes with $V$ and $D$ when the flow 
detaches. Finally, this independence of the lift on $V$ and $D$ explains why 
the lift on the half-circle only decreases after it reaches its maximum, 
positive value.

\begin{figure}[h]
\rotatebox{0}{\epsfig{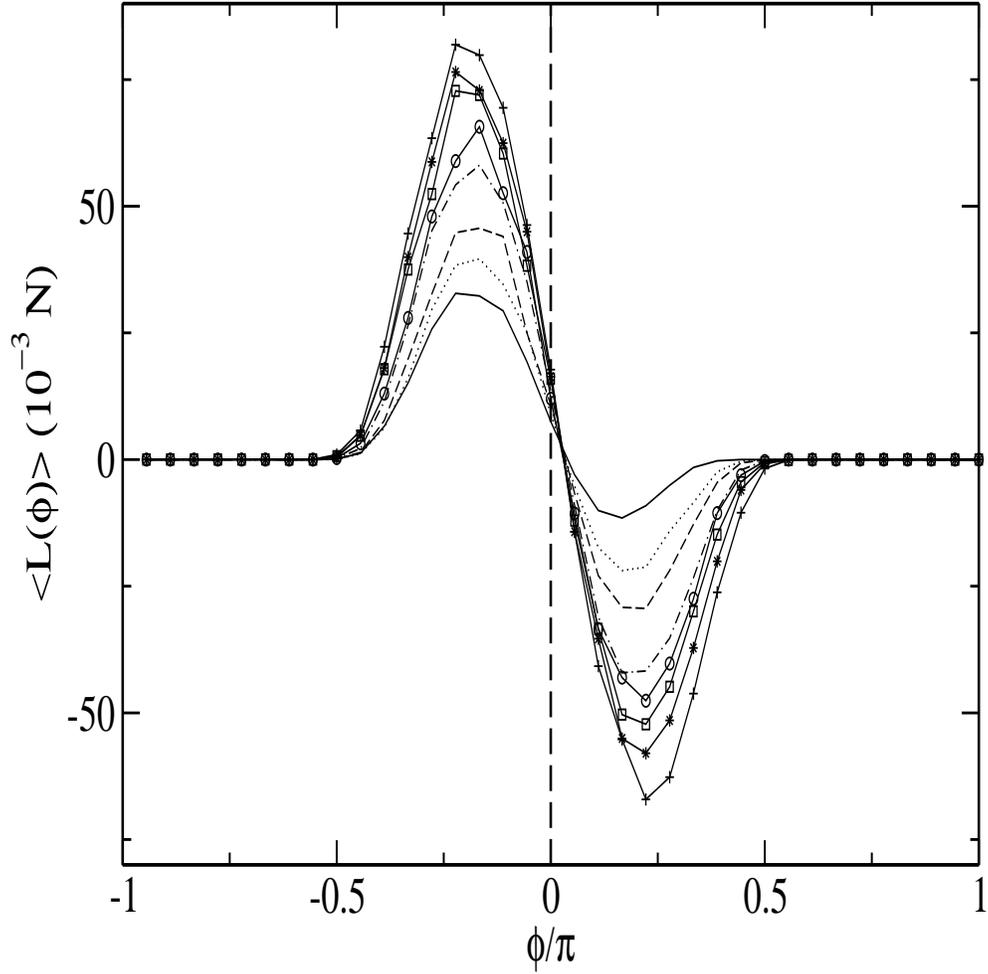}}
\caption{Lift profiles on the circle for $V=247$ cm/s. Depths (in cm): 
$3.75$ (solid), $7.50$ (dotted), $11.25$ (dashed), $18.75$ (dot dashed), 
$22.00$ (circles), $25.75$ (squares), $29.50$ (stars), $33.25$ (crosses).
\label{prof_lift_circle_2}}
\end{figure}

\begin{figure}[h]
\rotatebox{0}{\epsfig{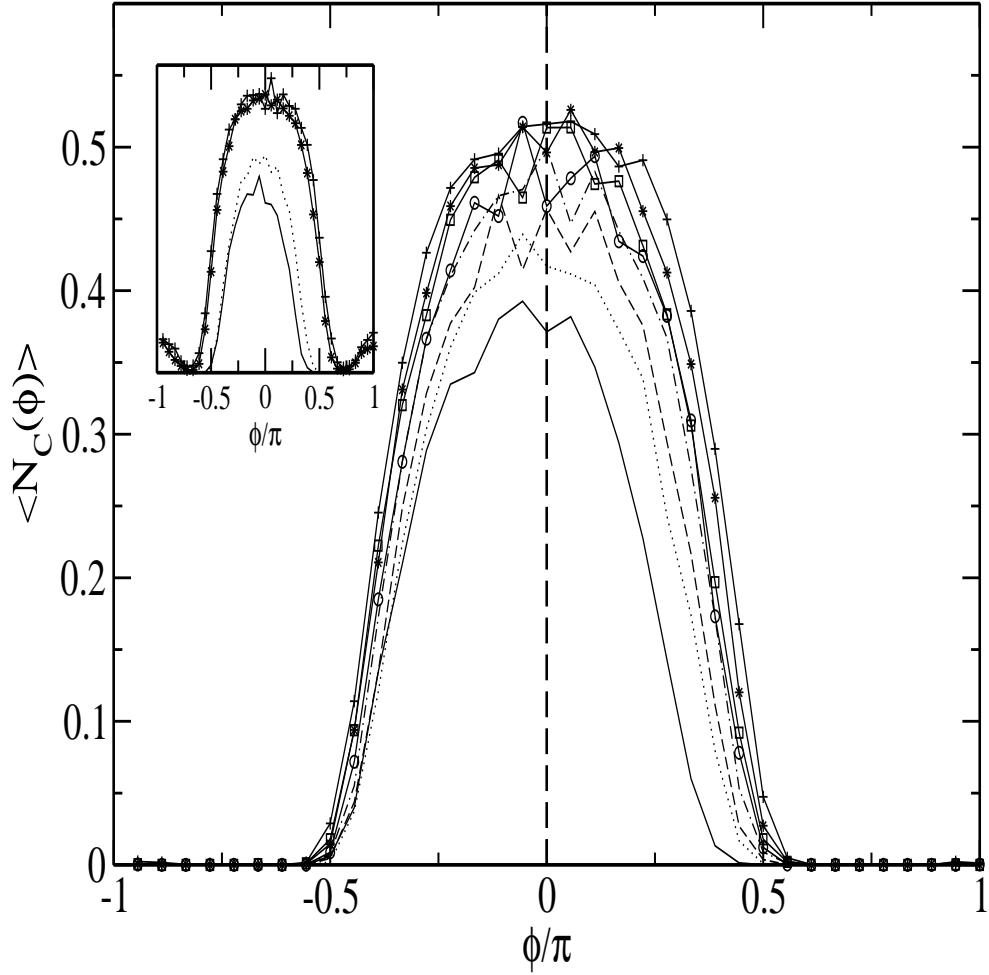}}
\caption{Contact number profiles for the circle at $V=247$ cm/s. 
Depths (in cm): $3.75$ (solid), $7.50$ (dotted), $11.25$ (dashed), $18.75$ 
(dot dashed), $22.00$ (circles), $25.75$ (squares), $29.50$ (stars), $33.25$ 
(crosses). Inset: $V=168$ cm/s, depths follows the same convention as main 
graph.
\label{prof_cont_circle_2}}
\end{figure}

Let us look at the lift and contact profiles as functions of depth in figs. 
\ref{prof_lift_circle_2} and \ref{prof_cont_circle_2}. The main feature of 
these plots is the overall increase of both quantities with depth and the 
familiar flow detachment, consistent with hydrostatic conditions. Our results 
also show that flow detachment can be suppressed at a given depth for a 
suitable speed. This is clear from the inset of fig. \ref{prof_lift_circle_2}, 
which shows contact profiles for intermediate speed. At low depths, contacts 
occur only on the leading side, $\phi=[-\pi/2,pi/2]$. For deep intruders, 
we find non-zero contact number all over the intruder surface. At even lower 
speed, we did not observe this phenomenon. Our results suggest we should go to 
a shallower position, at lower speeds, in order to see flow detachment. We 
cannot say if that would occur because for a very shallow intruder there will 
be only a few grain layers above the intruder, and the type of flow above it 
will so dilute that we will not a flow pattern similar to the ones we showed 
here.

\begin{figure}[h]
\rotatebox{0}{\epsfig{file=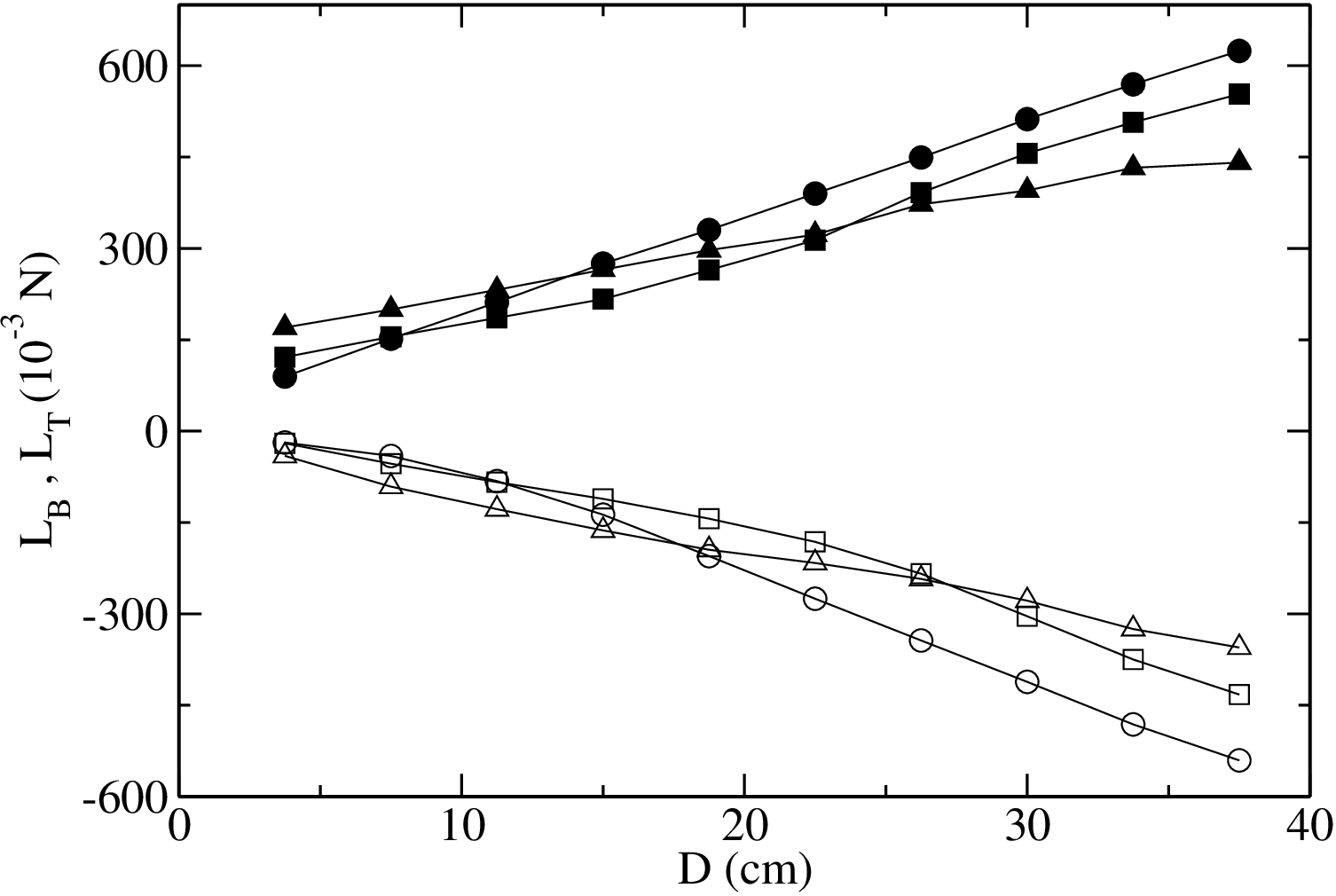,width=13.0cm,height=13.0cm}}
\caption{Lift on top (open) and bottom (filled symbols) of the circle as 
functions of depth. Speeds (in cm/s): $61.8$ (circles), $165$ (squares) and 
$247$ (triangles). Lines are guides to the eyes.
\label{lift_sides_circle_2}}
\end{figure}

\begin{figure}[h]
\rotatebox{0}{\epsfig{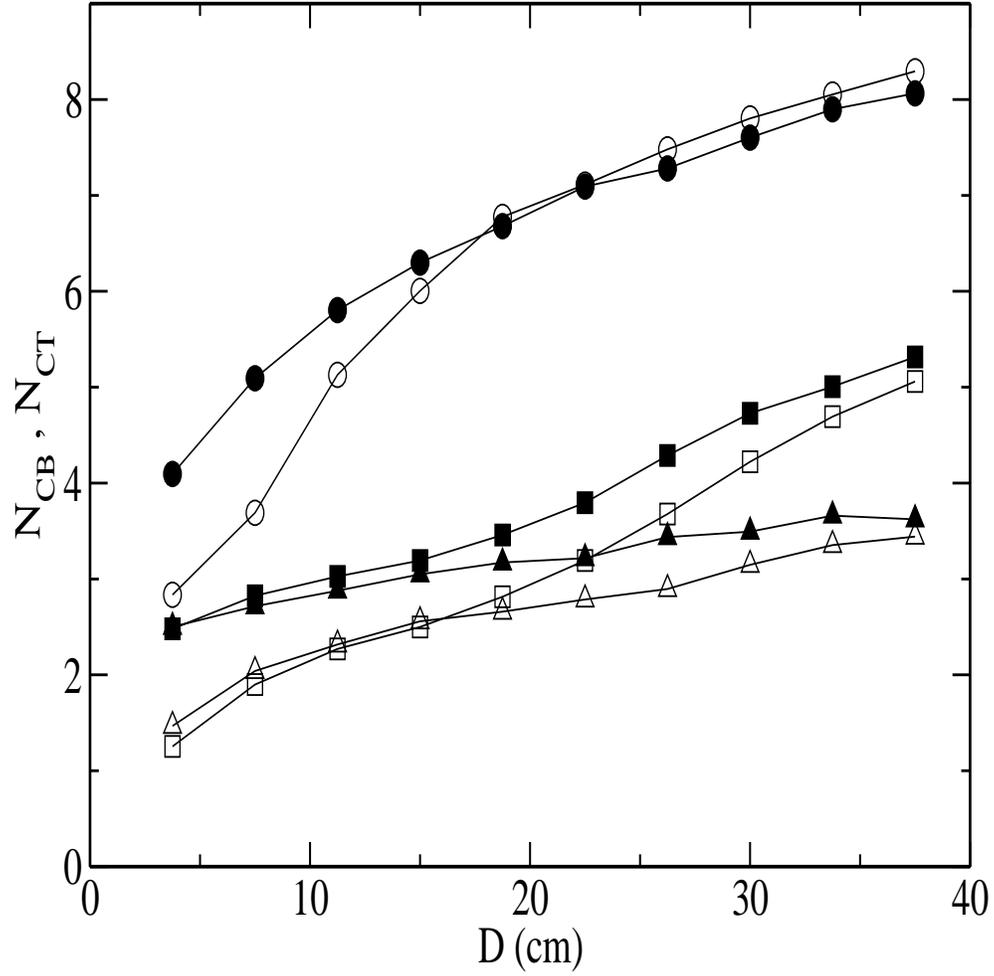}}
\caption{Contact numbers on top (open) and bottom (filled symbols) of the 
circle as functions of depth. Speeds (in cm/s): $61.8$ (circles), $165$ (squares) and $247$ (triangles). Lines are guides to the eyes.
\label{cont_sides_circle_2}}
\end{figure}

In figs. \ref{lift_sides_circle_2} and \ref{cont_sides_circle_2}, we show the 
lift and contact numbers on top and bottom sides of the circle as a function 
of depth. It is clear that the bottom lift is linear in $D$ for all speeds, and 
it changes slower with depth for high drag speeds. The top lift curves, 
however, do not suggest unique linear dependence between lift and depth, 
except for the $V=247$ cm/s curve. The other two show two distinct linear 
regimes, one at low and other at high speeds. Looking at the contact numbers, 
we see that the contact numbers have two distinct regimes with depth: a fast 
growth one at low $D$, and a slower variation one at higher depths. These 
results might explain the reason that the top lift changes behavior at some 
depth. This is consistent with previous speed results that showed that when the 
drag speed is below $V=227$ cm/s, the lift value is affected by changes in 
the contact numbers. It also explains why the initial fast growth of the top 
contact number with depth does not affect the lift value.

\begin{figure}[h]
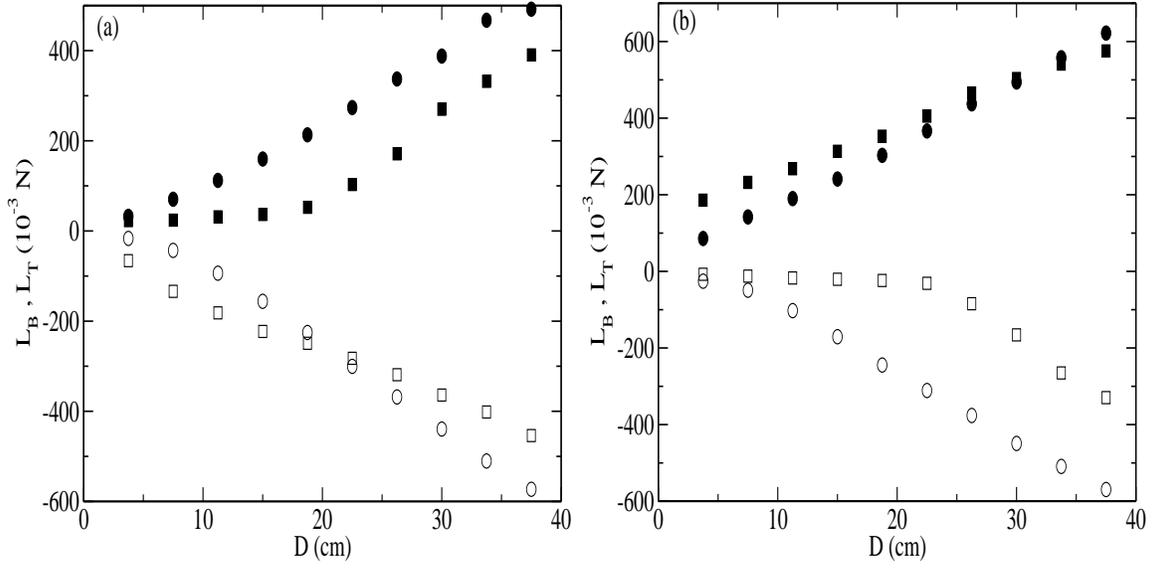

\centering
\begin{tabular}{cc}
\rotatebox{0}{\epsfig{file=Lift_Sides_P7_P8.eps,width=7.5cm,height=7.5cm}} &
\rotatebox{0}{\epsfig{file=Lift_Sides_P10_P11.eps,width=7.5cm,height=7.5cm}}\\
\end{tabular}
\caption{Lift on top (open) and bottom (filled symbols) of half-circle, panel 
(a), and inverted one, (b), as functions of depth. Speeds (in cm/s): 
$61.8$ (circles) and $247$ (squares). 
\label{lift_sides_both_2}}
\end{figure}

For the half-circles, we show, in fig. \ref{lift_sides_both_2}, top and 
bottom lifts as functions of depth. We see, again, the familiar depth 
independence of the lift on the flat side. In this case, it appears at high 
speed and low depth. Again, we can understand this result in view of the fact 
that flow detachment can be suppressed for suitable speed and depth. At low 
depth and high speeds, flow detaches. As the depth is increased, this 
phenomenon is decreased, and the net lift linearly increases with $D$, as 
seen in fig. \ref{lift_depth_both_1}. For other cases, the lift force increases 
approximately linear on depth.

\section{Conclusions\label{conclusions}}
We presented numerical results of the forces, drag and lift, on an intruder 
dragged 
horizontally through a hydrostatic granular packing, as functions of the drag 
speed, $V$, depth, $D$, and shape. We saw that the net drag increases 
quadratically with $V$ and linearly with $D$, and is larger for the circular 
intruder. The net lift, however, depends nonmonotonically on both $V$ and $D$ 
and is lower for the circle. We found that the sign of the net lift force 
depend on $V$ and $D$, as seen on the results for the half-circle. For the 
other two shapes, the net lift was always positive.

This complex behavior can be explained by detachment of the flow, i.e., 
decrease in the overall grain-intruder contact number. This effect decreases 
the forces on the intruder. Since this decrease is not the same on the upper 
and lower parts of the intruders with $V$, in some speed range the lift force 
on the lower side grows faster than the one on the upper side, which increases 
the net lift. The other effect is also seen, the lift on the upper part grows 
faster than the lift on the lower part, which is consistent with a decrease of 
the net lift. The effect of this phenomenon on the flat side of the 
half-circles is that all forces on this side are done on the leading point, 
since the flow is deviated at this point and barely touches the flat side at 
high $V$. Finally, we see that this effect can be suppressed by a high depth. 
This means that if an intruder is dragged at a speed where we see flow 
detachment, if this speed is keep and depth increased, the flow will touch the 
whole intruder surface as if it were dragged at a much slower speed. 

We also showed that the vertical forces on top and bottom sides of the 
intruders follow similar relationships with $V$ and $D$ as those for the net 
drag, $\left<R\right>$. This is to be expected, since all forces are contact 
grain-intruder interactions.

Future possibilities of these studies are on a detailed description of the 
forces within the packing as the intruder is dragged. If we know how the 
intruder will affect the forces in the flow, we will be able to predict much 
more accurately the beginning of flow detachment and why it does not occur in 
the same on both sides of an intruder. Also, a more detailed description of 
the lift on both top and bottom sides may lead to an empirical force law 
similar to those of the drag force obtained in \cite{Buchholtz98,Wass03}.

Finally, other shapes might show interesting results as well. For instance, 
we observed that if we drag an ellipse, we can also see a sign inversion for 
the lift as we saw in the half-circle case. In addition, this sign inversion 
seems to be dependent only on the angle of attack of the ellipse. Obviously, 
this effect depends also on eccentricity, speed and depth. If we understand 
this effect, we might be able to design more efficiently robots that move in 
granular environments, as those proposed in \cite{maladen11}, as well as 
understand better how animals locomote in sand \cite{maladen09}.

\section*{Acknowledgments} 
We are in debt with Daniel I. Goldman for his time in discussing and 
questioning from the point where the simulations were set up until the 
manuscript was finished. This work is financially supported by CAPES, CNPq and FAPESPA (brazilian agencies).


\section*{Figures}

\newpage 

\newpage

\newpage 

\newpage

\newpage

\newpage

\newpage

\newpage

\newpage

\newpage

\newpage

\newpage

\newpage

\newpage

\newpage

\vspace{2.0cm}

\newpage

\newpage

\vspace{2.0cm}

\vspace{2.0cm}

\vspace{2.0cm}

\end{document}